\documentclass[sigconf,letter, nonacm]{acmart}

\renewcommand\footnotetextcopyrightpermission[1]{} 
	\usepackage{graphicx} 
	\usepackage{balance}
    \usepackage{amssymb,amsmath}
    \usepackage{mathtools}
	\usepackage[utf8]{inputenc}

\usepackage{cleveref}
\crefname{section}{§}{§§}
\Crefname{section}{§}{§§}

	\usepackage{array}
    \usepackage{booktabs}
	\usepackage[utf8]{inputenc}
	\usepackage[english]{babel}
	\usepackage{comment}
	
	\usepackage{multirow, rotating, wasysym, url}
	\usepackage[ruled,linesnumbered,vlined]{algorithm2e}
    \usepackage{algorithmicx}  
	\usepackage{hyperref} 
	\usepackage{color}
    \usepackage{float}
    \usepackage{subcaption}
	\usepackage{soul}
	\usepackage{bbm}
	\usepackage[margin=0.05in]{caption}
	\usepackage{enumerate}
	\usepackage{framed}
	\usepackage{lipsum}
	\usepackage[normalem]{ulem}
	\usepackage{enumitem}
	\usepackage{makecell}

\newtheorem{lemma}{Lemma}
\newtheorem{theorem}{Theorem}

\graphicspath{{./DrawingFigures/},{./EvaluateFigures/}, {./GraphPPT/}} 

	\newcommand{\presec}{\vspace{-0.02in}}
	\newcommand{\postsec}{\vspace{-0.02in}}
	
	\newcommand{\presub}{\vspace{-0.03in}}
	\newcommand{\postsub}{\vspace{-0.02in}}

	\newcommand{\presubsub}{\vspace{-0.04in}}

	\newcommand{\prefig}{\vspace{-0.09in}}
	\newcommand{\postfig}{\vspace{-0.17in}}
	\newcommand{\presubfigcaption}{\vspace{-0.25in}}
	
	\newcommand{\prefigcaption}{\vspace{-0.2in}}

	\graphicspath{{./EvaluateFigures/}, {./GraphPPT/}, {./DrawingFigures/}}
	
	\newcommand{\ie}{\textit{i.e.}}
	\newcommand{\eg}{\textit{e.g.}}

	\newcommand{\pretheo}{\vspace{0cm}}
	\newcommand{\posttheo}{\vspace{0cm}}

	\definecolor{reder}{RGB}{255,0,0}

	\definecolor{zyd}{RGB}{0,0,255}
	\definecolor{gray}{rgb}{0.6,0.6,0.6}
	\definecolor{shadecolor}{rgb}{0.92,0.92,0.92}

	\newcommand{\ppp}{\noindent\textbf}

	\definecolor{yt}{RGB}{0,166,0}

	\definecolor{old}{RGB}{192,192,192}

	\definecolor{zyk}{RGB}{191, 63, 191}

	\definecolor{Chocolate}{RGB}{210,105,30}

	\newcommand{\aname}{ReliableSketch}

	\newcommand{\anamelong}{ReliableSketch}
	\newcommand{\goal}{SenseCtrlErr}
	
 \title[]{Approaching 100\% Confidence in Stream Summary through ReliableSketch}
    \author{Yuhan Wu}
    \affiliation{%
      \institution{Peking University}
      \city{Beijing}
      \country{China}
    }
    \email{yuhan.wu@pku.edu.cn}
    
    \author{Hanbo Wu}
    \affiliation{%
      \institution{Peking University}
      \city{Beijing}
      \country{China}
    }
    \email{wuhanbo@pku.edu.cn}
    
    \author{Xilai Liu}
    \affiliation{%
      \institution{Institute of Computing Technology, Chinese Academy of Sciences}
      \city{Beijing}
      \country{China}
    }
    \email{liuxilai20121013@gmail.com}
    
    \author{Yikai Zhao}
    \affiliation{%
      \institution{Peking University}
      \city{Beijing}
      \country{China}
    }
    \email{zyk@pku.edu.cn}
    
    \author{Tong Yang}
    \affiliation{%
      \institution{Peking University}
      \city{Beijing}
      \country{China}
    }
    \email{yangtong@pku.edu.cn}
    
    \author{Kaicheng Yang}
    \affiliation{%
      \institution{Peking University}
      \city{Beijing}
      \country{China}
    }
    \email{ykc@pku.edu.cn}
    
    \author{Sha Wang}
    \affiliation{%
      \institution{National University of Defense Technology}
      \city{Changsha}
      \country{China}
    }
    \email{ws0623zz@163.com}
    
    \author{Lihua Miao}
    \affiliation{%
      \institution{Huawei Technologies Co., Ltd.}
      \city{Shenzhen}
      \country{China}
    }
    \email{miaolihua4@huawei.com}
    
    \author{Gaogang Xie}
    \affiliation{%
      \institution{Computer Network Information Center, Chinese Academy of Sciences}
      \city{Beijing}
      \country{China}
    }
    \email{xie@cnic.cn}
    
\begin{document}


\begin{abstract}
    To approximate sums of values in key-value data streams, sketches are widely used in databases and networking systems. They offer high-confidence approximations for any given key while ensuring low time and space overhead. 
    While existing sketches are proficient in estimating individual keys, they struggle to maintain this high confidence across all keys collectively, an objective that is critically important in both algorithm theory and its practical applications.
    We propose ReliableSketch, the first to control the error of all keys to less than \(\Lambda\) with a small failure probability \(\Delta\), requiring only \(O(1 + \Delta\ln\ln(\frac{N}{\Lambda}))\) amortized time and \(O(\frac{N}{\Lambda} + \ln(\frac{1}{\Delta}))\) space. Furthermore, its simplicity makes it hardware-friendly, and we implement it on CPU servers, FPGAs, and programmable switches. Our experiments show that under the same small space, ReliableSketch not only keeps all keys' errors below \(\Lambda\) but also achieves near-optimal throughput, outperforming competitors with thousands of uncontrolled estimations. We have made our source code publicly available.
        
\end{abstract}


	\maketitle

\presec
\section{Introduction} \postsec
\label{sec:intro}

In data stream processing, stream summary \cite{cmsketch} is a simple but challenging problem: within a stream of key-value pairs, query a key ``$e$'' for its value sum $f(e)$ --- the sum of all values associated with that key.
The problem is typically addressed by ``sketches'' \cite{melissourgossingle, li2022stingy, zhao2023panakos}, a kind of approximate algorithm that can answer an estimated sum \(\hat{f}(e)\) with small time and space consumption.
In terms of accuracy, existing sketches ensure that the absolute error of \(\hat{f}(e)\) is less than \(\Lambda\) with a high probability \(1-\delta\). This can be formally expressed as:
\[\text{For arbitrary key } e, \Pr\left[\left|\hat{f}(e) - f(e)\right| \leqslant \Lambda\right] \geqslant 1-\delta,\]
where there are two critical parameters: the error tolerance \(\Lambda\), under which the absolute error is considered controllable, and \(\hat{f}(e)\) is deemed sufficiently accurate; otherwise, the key \(e\) is referred to as an outlier. The \textit{individual Confidence Level (CL)}, \(1-\delta\), represents the lower bound probability that \(\hat{f}(e)\) is sufficiently accurate.

Existing sketches, effective for individual queries, struggle with accurately answering multiple queries at once.
When \(N\) keys are queried collectively, \textit{the overall CL}, denoted as \(1-\Delta\), that all answers are sufficiently accurate equals \((1-\delta)^N\). The overall CL rapidly decreases as \(N\) increases: from \((1-\delta) = 95\%\) for a single key to 90.25\% for two keys, and further diminishes to just 1\% for 90 keys.
Furthermore, when all keys are queried collectively, as in a million-key scenario, a significant absolute number---about the \(\delta\)-fraction of these keys---are expected to be outliers.
These outliers, which users cannot distinguish from other keys, undermine confidence in the results' reliability and poses real-world challenges. For example, in network devices, sketches are used to identify if a key is frequent (if its value is large enough). In a dataset with 1 million infrequent and 1,000 frequent keys, even with a 99\% individual CL, approximately 10,000 infrequent keys might be wrongly labeled as frequent, leading to a high false positive rate of 90.9\%. Such misidentification can cause serious issues in network applications, such as placing critical control signals into low-priority queues, which can result in the loss of these signals during network congestion.

Our target problem is to accurately answer an unlimited number of queries collectively, with a negligible failure probability \(\Delta\).
This can be formally stated as: \[\Pr\left[\forall \textit{ key } e, \left|\hat{f}(e)-f(e)\right|\leqslant\Lambda\right]\geqslant 1-\Delta,\]

\begin{table*}[h]
\centering
\begin{tabular}{|c|l|l|l|l|}
\hline
& \begin{tabular}[c]{@{}l@{}}\textbf{Counter-based }\\ \textbf{(L1-Norm)}\end{tabular} & \begin{tabular}[c]{@{}l@{}}\textbf{Counter-based }\\ \textbf{(L2-Norm)}\end{tabular}  & \textbf{Heap-based} & \textbf{Reliable Sketch (Ours)} \\
\hline
Overall Confidence & \begin{tabular}[c]{@{}l@{}}Moderate: \\ \((1 -\delta)^N\)\end{tabular} & \begin{tabular}[c]{@{}l@{}}Moderate: \\ \((1 -\delta)^N\)\end{tabular} & \begin{tabular}[c]{@{}l@{}}Optimal: \\ 100\%\end{tabular} & \begin{tabular}[c]{@{}l@{}}Near Optimal: \\ \(1 - \Delta\)\end{tabular} \\
\hline
Speed & \begin{tabular}[c]{@{}l@{}}Moderate: \\ \(O(\ln(\frac{1}{\delta}))\)\end{tabular} & \begin{tabular}[c]{@{}l@{}}Moderate: \\ \(O(\ln(\frac{1}{\delta}))\)\end{tabular} & \begin{tabular}[c]{@{}l@{}}Low: \\ \(O(\ln(\frac{N}{\Lambda}))\)\end{tabular} & \begin{tabular}[c]{@{}l@{}}High: \\ \(O(1+\Delta\ln\ln(\frac{N}{\Lambda}))\)\end{tabular} \\
\hline
Space & \begin{tabular}[c]{@{}l@{}}Low: \\ \(O(\frac{N}{\Lambda} \times \ln(\frac{1}{\delta}))\)\end{tabular} & \begin{tabular}[c]{@{}l@{}}Low: \\ \(O(\frac{N_2^2}{\Lambda^2} \times \ln(\frac{1}{\delta}))\)\end{tabular} & \begin{tabular}[c]{@{}l@{}}Optimal: \\ \(O(\frac{N}{\Lambda})\)\end{tabular} & \begin{tabular}[c]{@{}l@{}}Near Optimal: \\ \(O(\frac{N}{\Lambda} + \ln(\frac{1}{\Delta}))\)\end{tabular} \\
\hline
Compatibility & High & High & Low & High \\
\hline
\end{tabular}
\caption{Comparative Analysis of Counter-based, Heap-based, and Reliable Sketches: Here, \(N=\sum f(e)\), \(N_2=\sqrt{\sum (f(e))^2}\), and \(\Lambda\) is the error tolerance. Additionally, \(\delta\) and \(\Delta\) represent the probabilities of failure for individual and all keys, respectively.}

\label{tab:sketch_comparison_transposed}
\end{table*}

As Table \ref{tab:sketch_comparison_transposed} illustrates, existing sketches, both counter-based and heap-based, can hardly address our target problem with small time and space consumption.
Counter-based sketches \cite{cmsketch, countsketch, melissourgossingle, basat2021salsa}, which only record counters, increase confidence by repeating experiments and creating multiple sub-sketch copies. To achieve the individual CL of \(1-\delta\), they create \(\ln(\frac{1}{\delta})\) copies, which results in a time and space cost multiplied by \(\ln(\frac{1}{\delta})\). 
For keys that can potentially reach a number up to $N$ ($N:= \sum f(e)$), accurately estimating all keys requires setting $\delta$ to a very small fraction of $\frac{\Delta}{N}$. This results in a significant increase in time and space costs.
Counter-based sketches are divided into two types based on complexity: those using the L1 norm (e.g., CM \cite{cmsketch}, CU \cite{estan2002cusketch}, Elastic \cite{yang2018elastic}) and the L2 norm (e.g., Count \cite{countsketch}, UnivMon \cite{liu2016UnivMon}, Nitro \cite{liu2019nitrosketch}). Given the challenge in assuming data characteristics, these types are generally not directly comparable. Our research focuses on optimizing L1 norm-based sketches.
Heap-based sketches, such as Space Saving (SS) \cite{spacesaving} and Frequent \cite{frequent1}, are more adept at dealing with outliers but suffer from slower data insertion due to their logarithmic time complexity $\left(O(\ln(\frac{N}{\Lambda}))\right)$ heap structures. Additionally, they face compatibility challenges with high-speed hardware like FPGAs.

To address the target problem, we propose the ReliableSketch with versatile theoretical and practical advantages:
\begin{itemize}[leftmargin=0.1in]
    \item Confidence: ReliableSketch guarantees that the error for all keys is controllable with an overall CL of \(1-\Delta\), where \(\Delta\) is a small quantity that can be easily reduced to below \(10^{-10}\). This ensures that not a single outlier will occur even after many years of the algorithm's operation. Here, "controllable" refers to keeping the error less than \(\Lambda\).
    \item Speed: Our time complexity is lower than existing solutions. The amortized time cost for inserting each key-value pair is only \(O(1 + \delta\ln\ln(\frac{N}{\Lambda}))\). In practice, \(\ln\ln(\frac{N}{\Lambda})\) is generally much smaller than \(\frac{1}{\delta}\), making the time complexity effectively O(1) most of the time.
    \item Space: Our space complexity is \(O(\frac{N}{\Lambda} + \ln(\frac{1}{\delta}))\), which is efficient because the result is additive and \(\frac{N}{\Lambda}\) is generally much greater than \(\ln(\frac{1}{\delta})\). We can easily set \(\delta\) to be very small without worrying about increased space and time overhead.
    \item Compatibility with High-Speed Hardware: Our design is friendly to high-performance hardware architectures like FPGA, ASIC, and Tofino, adhering to the programming constraints of pipeline architecture. Our data structure does not require pointers, sorting, or dynamic memory allocation.
\end{itemize}
Compared to counter-based sketches, we have improved in terms of confidence, speed, and space complexity. We ensure high overall confidence for all keys collectively, reducing the amortized insertion complexity and transforming the space cost from a multiplicative \(O(\frac{N}{\Lambda} \times \ln(\frac{1}{\delta}))\) to an additive one. Compared to heap-based sketches, we have made significant improvements in speed, optimizing the non-parallel \(O(\ln(\frac{N}{\Lambda}))\) time to amortized O(1), while achieving nearly the same level of confidence and space efficiency.

Our main strategy is to sense errors in all keys in real-time, controlling those with larger errors to prevent any from becoming outliers.
There are two key challenges involved: how to measure errors and how to control them. In response, we address them by two key techniques respectively: the Error-Sensible bucket and the Double Exponential Control.

Challenge 1: Measure Errors. Existing sketches, like the count-min sketch, do not know the error associated with a key because they mix the values of different keys in the same counter upon hash collisions.
We harness an often-undervalued feature of the widely adopted election technology \cite{tang2019mvsketch, frequent1, yang2018elastic}, particularly emphasizing the role of the negative vote. 
In elections, negative votes provide an ideal way to observe the hash collision and sense the error for each key during estimation. We replace the counter with an Error-Sensible bucket that contains an election mechanism, enabling real-time observation of each key's error.

Challenge 2: Controlling Errors. Existing sketches control errors by constructing $d$ identical sub-tables for repeated experiments. This approach not only incurs significant overhead, as repeating the experiment $d$ times requires $d$ times the resources in terms of time and space, but it can only eliminate $O(e^d)$ outliers. Our Double Exponential Control technique not only eliminates $O(e^{(e^d)})$ outliers (with d=8, $e^{(e^d)} > 10^{1294}$) among all keys using $d$ sub-tables, but also maintains a steady time and space cost, which does not increase linearly with the growth of $d$.

Our key contributions are as follows:
\begin{itemize}[leftmargin=0.2in]
    \item We devise \textbf{ReliableSketch} to  accurately answer the queries for all keys collectively, with a negligible failure probability $\Delta$.
    \item We theoretically prove that \anamelong{} outperforms state-of-the-art in both space and time complexity.
    \item We implement \anamelong{} on multiple platforms including CPU, FPGA, and Programmable Switch. Under the same small memory, ReliableSketch not only eliminates outliers, but also achieves near the best throughput, while its competitors have thousands of outliers.
\end{itemize}

	\presec
\section{Background and Motivation} \postsec
\label{sec:relate}
In this section, we start with the problem definition and then discuss the limitations of existing sketches, using the typical CM Sketch as an example.

\presub
\subsection{Problem Definition} \postsub
\label{subsec:related:problem}
\label{subsec:related:usecase}

\ppp{Stream Summary Problem \cite{cmsketch}.} In a key-value data stream \( S=\{\langle e_1, v_1\rangle, \langle e_2, v_2\rangle, \dots\} \), the sketch algorithm processes each pair (or data item) in real-time. At any moment, for a user query about a specific key \( e \), the sketch can rapidly estimate the aggregate sum of values for all pairs containing \( e \). The actual sum and the sketch's estimated sum for a key \( e \) are denoted as \( f(e) \) and \( \hat{f}(e) \), respectively. Within a predefined error tolerance \( \Lambda \), a key \( e \) whose estimation error exceeds \( \Lambda \), expressed as \( \left|\hat{f}(e) - f(e)\right| > \Lambda \), is defined as an outlier.

We aim at accurately answering all queries collectively and eliminating outliers, under user-specified hyperparameters \( \Lambda \) (the error tolerance) and \( \Delta \) (upper limit of failure probability): \[\Pr\left[\forall \textit{ key } e, \left|\hat{f}(e)-f(e)\right|\leqslant\Lambda\right]\geqslant 1-\Delta.\]

\presub
\subsection{Limitation of Existing Solutions} \postsub

Here, for readers unfamiliar with sketches, we begin with the simplest CM sketch to explain why existing sketches struggle to eliminate outliers. A detailed complexity analysis has already been discussed in Table \ref{sec:intro} and \cref{tab:sketch_comparison_transposed}, and will not be reiterated.

The CM sketch is a prime example of the design philosophy behind most counter-based sketches. It comprises \(d\) arrays, \(A_i[.]\), each containing \(w\) counters. For a key \(e\), CM selects \(d\) \textit{mapped counters} using independent hashing. The \(i\)-th mapped counter is \(A_i[h_i(e)]\), where \(h_i(\cdot)\) is the hash function. When an item arrives with key \(e\) and a positive value \(v\), CM increments these mapped counters by \(v\). To query the value sum of \(e\), CM reports the smallest counter among \(e\)'s mapped counters as the estimate. However, when other keys collide with the same counter as \(e\) (a collision), they add their value to the counter, causing an error. The key insight is that the smallest mapped counter is the most accurate due to the fewest collisions.

However, even the minimal counter, with the least error, can have significant inaccuracies when every mapped counter experiences severe hash collisions. Thus, CM and other counter-based sketches can only maintain a high confidence level for a single query. When querying a large number of keys, it's not guaranteed that each key's error will be small. Similarly, other counter-based sketches, including Count, CU, Univmon, and others, share this design limitation. The complexity of counter-based sketches is based on the L1 norm \(N=\sum f(e)\) and the L2 norm \(N_2=\sqrt{\sum (f(e))^2}\). Since the relative sizes of \(N\) and \(N_2\) depend on dataset characteristics, the complexity of these two types of sketches cannot be directly compared. Our research focuses on optimizing L1 norm-based sketches, while solutions based on the L2 norm's complexity for our target problem are left for future work..

Heap-based sketches, like Space Saving and Frequent, use heap structures to maintain high-frequency elements but suffer from slower data insertion due to their logarithmic time complexity ( \(\ln(\frac{N}{\Lambda})\)). This complexity cannot be accelerated through parallelization. Additionally, the pointer operations they require become an obstacle in implementing them on high-speed hardware platforms like FPGA programmable switches. Only when \(v=1\) can these heap structures be implemented with $O(1)$ complexity using linked lists, but we aim to address a broader range of stream summary problems where \(v\) is not equal to 1.

\newcommand\floor[1]{{\lfloor #1 \rfloor}}

\newcommand{\hit}{YES}
\newcommand{\miss}{NO}

\section{Reliable Sketch} \postsec
\label{SEC:Algorithm}

In this section, we present \anamelong{}. We start from a new alternative to the counter of counter-based sketches, termed an 'Error-Sensible bucket'. This allows every basic counting unit within \anamelong{} to perceive the magnitude of its error, as discussed in \cref{subsec:alg:onebucket}. Then, we demonstrate how to organize these buckets and set appropriate thresholds to control the error of every key within the user-defined threshold, as discussed in \cref{subsec:alg:formal:tech2}.

\begin{figure}[tbp]
    \centering  
    \includegraphics[width=0.96\linewidth]{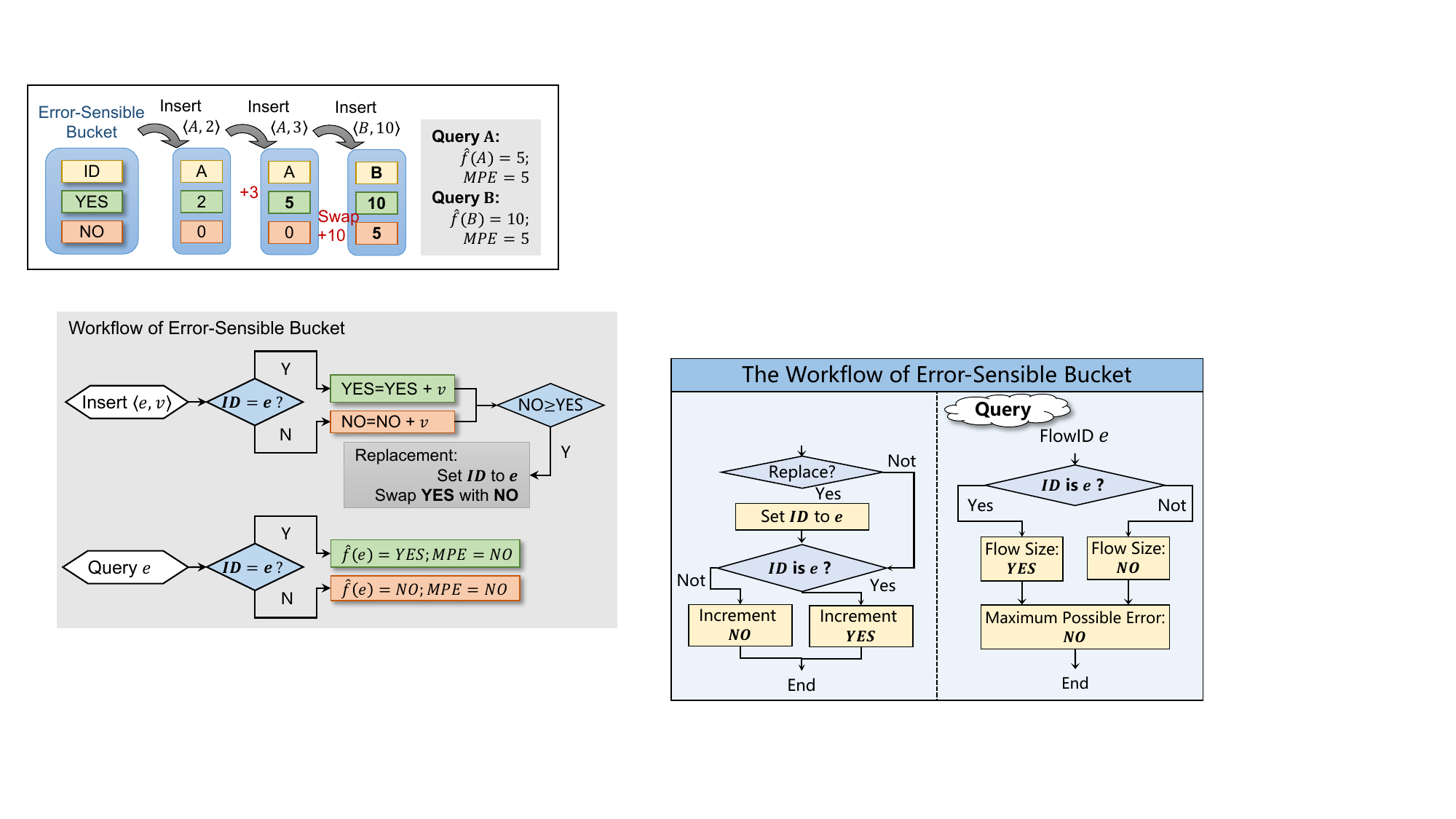}
    \caption{The workflow of the Error-Sensible Bucket, including insertion and querying processes.}
    \label{fig:operations:bucket:workflow}
    \postfig
\end{figure}

\begin{figure}[tbp]
    \centering  
    \includegraphics[width=0.96\linewidth]{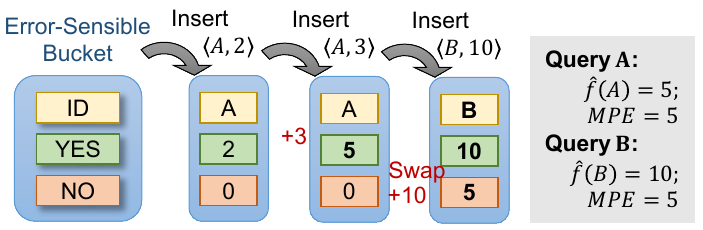}
    \caption{An example of how an Error-Sensible Bucket works: starting from empty, sequentially inserting three items followed by two queries.}
    \label{fig:operations:bucket:example}
    \postfig
\end{figure}

\presub
\subsection{Basic Unit: the Error-Sensible Bucket } \postsub
\label{subsec:alg:onebucket}

The basic unit is the smallest cell in a sketch that performs the counting operation. In counter-based sketches, the basic unit is a standard counter. In our \anamelong{}, the basic unit is the ``Error-Sensible Bucket'' structure, which actively perceives the extent of hash collisions and reports the Maximum Possible Error (MPE).
We demonstrate the workflow and a practical example in Figure \ref{fig:operations:bucket:workflow} and \ref{fig:operations:bucket:example}, respectively.

Formally, the bucket supports two operations: (1) Insert a key-value pair \(\langle e, v\rangle\); (2) Query the value sum of a key \(e\). Upon querying, the bucket returns two results, the estimated frequency \(\hat{f}(e)\) along with its MPE, satisfying \(\hat{f}(e) \in [f(e), f(e)+MPE]\).

\ppp{Structure:}
The bucket has three fields: one ID field recording a candidate key, and two counters recording the positive and negative votes for the candidate, denoted as \(ID\), \(YES\), and \(NO\), respectively. Initially, \(ID\) is null and both counters are set to 0.

\ppp{Insert.}
When inserting an item \(\langle e, v\rangle\), there are two phases: voting and replacement. If the newly arrived \(e\) is the same as \(ID\), a positive vote is cast, setting \(YES\) to \(YES + v\); otherwise, a negative vote is cast, making \(NO = NO + v\). If the positive votes are less than or equal to the negative votes, a replacement occurs: \(ID\) is set to \(e\), and the values of \(YES\) and \(NO\) are swapped.

\ppp{Query.}
When querying the value sum of \(e\), we first check if the recorded \(ID\) matches \(e\). If it does, indicating \(e\) is the current candidate, we use \(YES\) to estimate its value sum. It can be proven inductively that \(YES\) is greater than or equal to \(f(e)\), with its MPE being \(NO\), i.e., \(f(e) \in [YES - NO, YES]\). On the other hand, if \(ID\) is not equal to \(e\), we use \(NO\) as the estimate, which is always greater than \(f(e)\). The MPE remains \(NO\).

\ppp{Discussion---Correctness.}
The correctness of the above query can be directly proven through induction. Here, we discuss a more intuitive understanding. Consider a specific key \(e_0\) and its value sum \(f(e_0)\). Items with \(e_0\) in the data stream may cast either positive or negative votes, thus \(YES + NO\) is always greater than or equal to \(f(e_0)\). But why, when \(ID = e_0\), is \(YES + NO\) reduced to \(YES\), ensuring it's greater than or equal to \(f(e_0)\), and when \(ID \neq e_0\), \(NO\) is greater than or equal to \(f(e_0)\)?

The key to this proof lies in recognizing that \(NO\) records the "collision amount." Consider two scenarios where \(NO\) increases: (1) Without replacement, item \(\langle a, v \rangle\) increases \(NO\) by \(v\), with \(ID\) recorded as \(b\). This results in a collision amount \(v\) between \(a\) and \(b\), which are distinct. (2) With replacement, consider a bucket \(\langle ID = b, YES = y_0, NO = n_0 \rangle\) before replacement. Item \(\langle a, v \rangle\) increases \(NO\) by \(v\) and causes a swap of \(YES\) and \(NO\). The actual increase in \(NO\) is the difference between \(YES\) and \(NO\) before the swap, \(delta(NO) = y_0 - n_0\). This is deemed a collision between \(a\) and \(b\), who are distinct. Overall, \(NO\) represents the collision amount between two different keys, and any unit of value is part of at most one collision, not multiple.

Based on the collision amount, the rationality can be explained easily. A bucket is inserted with a total value of \(YES + NO\), where \(NO\) represents collisions. Since the recorded \(ID = e_0\) cannot collide with itself, at least \(NO\) values must be from keys other than \(e_0\), hence \(YES \geq f(e_0)\). Additionally, all increases in \(YES - NO\) are caused by the insertion of item \(e_0\), with a value sum of at least \(YES - NO\). Therefore, \(f(e_0) \in [YES - NO, YES]\). Conversely, when \(ID = e_1 \neq e_0\), deducting the part belonging to \(e_1\) from the total \(YES + NO\) leaves \(2 \times NO\). Since collisions always occur between different keys, at most \(NO\) units of value in the remaining \(2 \times NO\) can belong to the same key, thus \(f(e_0) \in [0, NO]\).

\begin{figure}[tbp]
    \centering  
    \includegraphics[width=0.98\linewidth]{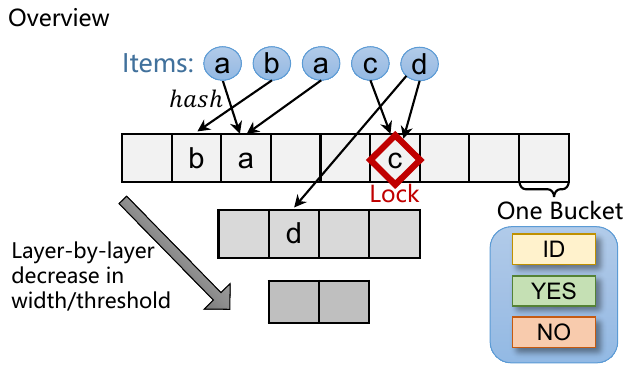}
    \vspace{-0.15in}
    \caption{An Overview of \anamelong{}.}
    \vspace{-0.1in}
    \label{fig:algo:structure}
    \postfig
\end{figure}

\presub
\subsection{Formal \aname{}} \postsub
\label{subsec:alg:formal:tech2}

In this part, we propose how to organize Error-Sensible buckets and integrate them into a \anamelong{} that can control the error of all keys. Our key idea is to lock the error of a bucket when it reaches a critical threshold, diverting any further error-increasing insertions to the next layer. We first introduce the ``lock'' mechanism, then describe the data structure of \anamelong{}, as well as its insertion and querying operations. 

\ppp{Lock Mechanism.}
When the Maximum Possible Error (MPE) of a bucket reaches a threshold, i.e., \(B.NO = \lambda \And B.YES>B.NO\), the bucket must be locked to halt the growth of MPE. Upon being locked, only two types of insertions are permitted, neither of which will increase the MPE (\(B.NO\)):

\begin{itemize}[leftmargin=*]
    \item If \(B.ID = e\), it only increments \(B.\hit\).
    \item If \(B.YES = B.NO\), a replacement occurs, and this also only increments \(B.\hit\).
\end{itemize}

In other scenarios, items cannot be inserted into the locked bucket. These items, at a higher risk of losing control, are redirected to other buckets for insertion.

\ppp{Data structure.}
As depicted in Figure \ref{fig:algo:structure}, \anamelong{} is composed of \(d\) layers, with each layer, indexed by \(i\), containing \(w_i\) Error-Sensible buckets, where the width \(w_i\) diminishes progressively with an increase in \(i\). In each layer, the \(j\)-th bucket is denoted as \(B_i[j]\).
Each layer is assigned a specific threshold, referred to as \(\lambda_i\), used to determine when to lock a bucket in the $i$-th layer. 
The \(\lambda_i\) values also decrease with increasing \(i\), and their cumulative sum does not surpass the user-defined error threshold, i.e., \(\sum_i \lambda_i \leqslant \Lambda\). 
Furthermore, each layer utilizes an independent hash function \(h_i(\cdot)\), which uniformly maps each key to a single bucket within the respective layer. 

\ppp{Parameter Configurations:} 
\anamelong{} performs best when both \(w_i\) and \(\lambda_i\) are set to \textbf{decrease exponentially}, which is our Key Technique II (Double Exponential Control). Modifying either parameter to follow an arithmetic sequence would thoroughly undermine the complexity of \anamelong{}. Discussions on its insights and novelty can be found at the end of 
\cref{subsec:alg:formal:tech2}. Practically, we set $w_i=\left\lceil\frac{W (R_{w}-1)}{R_{w}^i}\right\rceil$ and $\lambda_i = \left\lfloor\frac{\Lambda (R_\lambda-1)}{R_\lambda^i}\right\rfloor$, where $W$ denote the total number of buckets.
In our proofs (Theorem \ref{theo:main}), we set $W=\frac{4(R_w R_\lambda)^{6}}{(R_w-1)(R_\lambda-1)}\cdot \frac{N}{\Lambda}$, which is with large constant.
But based on our experiment results, we recommend to set $W=\frac{(R_w R_\lambda)^{2}}{(R_w-1)(R_\lambda-1)}\cdot \frac{N}{\Lambda}$, $R_w\in[2, 10]$, $R_l\in[2, 10]$, and $d\geqslant 7$.
When then memory size is given without given $\Lambda$, we set $\Lambda=\frac{(R_w R_\lambda)^{2}}{(R_w-1)(R_\lambda-1)}\cdot \frac{N}{W}$.



\begin{algorithm}[!h]
\renewcommand\baselinestretch{0.9}\selectfont
    \SetAlgoNoLine
    \caption{Insert Operation.}
    \label{alg:pseudo:Insertion:all:Weighted}
    \SetKwFunction{FMain}{Insert}
    \SetKwProg{Fn}{Procedure}{:}{}
    \Fn{\FMain{$\langle e, v\rangle$}}{
    
        \For{Layer $i=1,2,\dots, d$}{
            $j\gets h_i(e)$  
            
            \If{$B_i[j].ID=e$}{

                $B_i[j].YES\gets B_i[j].YES + v$
                
                \textbf{Return} 
            }
            \eIf{$B_i[j].NO+v>\lambda_i$ \textbf{and} $B_i[j].YES>\lambda_i$}{
                \Comment{Lock triggered}
            
                $B_i[j].NO\gets \lambda_i$
                
                $v\gets v-(\lambda_i-B_i[j].NO)$
                
                \textbf{Continue} \Comment{Continue to next layers}
            }{
                $B_i[j].NO\gets B_i[j].NO+v$
                
                \If{$B_i[j].NO\geqslant B_i[j].YES$}{
                    $B_i[j].ID\gets e$

                    Swap($B_i[j].NO$, $B_i[j].YES$)
                }
                \textbf{Return} 
            }
        }
    }
\end{algorithm}

\ppp{Insert Operation for Item \(\langle e, v\rangle\) (Algorithm \ref{alg:pseudo:Insertion:all:Weighted}).}
The insertion into \anamelong{} is a layer-wise process, starting at the first layer and continuing until the value $v$ is fully inserted. The operation, which may not involve all layers, includes these four steps in each layer:
\begin{enumerate}
    \item \textbf{Locating a Bucket (Line 3):} Utilizing the hash function \(h_i()\) of the current layer \(i\), we locate the \(j\)-th bucket, \(B_i[j]\), where \(j = h_i(e)\). We aim to insert \(\langle e, v\rangle\) specifically into this bucket, without considering other buckets in the same layer.
    
    \item \textbf{Handling Matching ID (Line 4-7):} If \(B_i[j].ID\) equals \(e\), we increment \(YES\) by \(v\) and finish the insertion without moving to further layers.

    \item \textbf{Triggering Lock (Line 8-12):} This step follows if the ID doesn't match. Before increasing \(NO\), we check if adding \(v\) triggers the layer threshold \(\lambda_i\). The lock activates when \(B_i[j].NO + v > \lambda_i\) (indicating certain lock activation without replacement) and when \(B_i[j].YES > \lambda_i\) (signifying lock activation upon replacement). On triggering the lock, only a portion of \(\langle e, v\rangle\) can be accommodated in \(B_i[j]\), equal to the difference \(\lambda_i - B_i[j].NO\). The excess value, \(v - (\lambda_i - B_i[j].NO)\), is reserved for insertion into subsequent layers (Line 12). Consequently, \(B_i[j].NO\) is adjusted to \(\lambda_i\).

    \item \textbf{Adjusting NO and Checking for Replacement (Line 14-19):} If this step is reached, it means a negative vote is cast, and the lock would not be activated. \(B_i[j].NO\) is incremented by \(v\). Then, compare \(B_i[j].NO\) with \(B_i[j].YES\) to determine if a replacement occurs. If it does, perform the replacement. The insertion is finished, and no further layers are visited.
\end{enumerate}

If, by the end of the final layer, there remains value that has not been inserted, we consider the insertion operation to have failed. Once insertion failure occurs, we cannot guarantee zero outliers. Fortunately, through our design and theoretical proofs, we have shown that the probability of such an failure is extremely low. For those still concerned about this scenario, refer to \cref{extension:emergency} for emergency solutions.

\begin{algorithm}[!h]
\renewcommand\baselinestretch{0.9}\selectfont
    \SetAlgoNoLine
    \caption{Query Operation.}
    \label{alg:pseudo:query:all}
    \SetKwFunction{FMain}{Query}
    \SetKwProg{Fn}{Function}{:}{}
    \Fn{\FMain{$e$}}{
        
        $ \hat{f}(e)\gets 0$  \Comment{Estimator}
                
        $ MPE\gets 0$ \Comment{Maximum Possible Error}
    
        \For{Layer $i=1,2,\dots, d$}{
            $j\gets h_i(e)$  
            
            \eIf{$B_i[j].ID=e$}{
                $\hat{f}(e)\gets \hat{f}(e) + B_i[j].YES$
                
            }{
                $\hat{f}(e)\gets \hat{f}(e) + B_i[j].NO$
                
            }

            $ MPE\gets MPE + B_i[j].NO$

            \If{ ( $B_i[j].NO  < \lambda_i$ \textbf{or} $B_i[j].YES = B_i[j].NO$ \textbf{or} $B_i[j].ID=e$)}{
                \textbf{break} \Comment{Stop collecting value.}
            }
        }
        
        \Return{$\langle \hat{f}(e), MPE\rangle$}
    }
\end{algorithm}

\ppp{Query Operation for Item \(e\) (Algorithm \ref{alg:pseudo:query:all}).}
In \anamelong{}, each query reports \(\hat{f}(e)\) along with its Maximum Possible Error (MPE). The query operation is similar to insertion, requiring a layer-wise process to gather results layer by layer, stopping as soon as there's sufficient reason to do so (which usually happens quickly).
Beginning from the first layer, we sequentially access the hashed bucket \(B_i[j], j=h_i(e)\) in each layer. If \(B_i[j].ID\) equals \(e\), we add \(B_i[j].YES\) to \(\hat{f}(e)\); otherwise, we add \(B_i[j].NO\). For MPE, we always add \(B_i[j].NO\).
The query can be finished without accessing subsequent layers if any of the following conditions are met, as each indicates that \(e\) has not been inserted into subsequent layers: (1) \(B_i[j].NO < \lambda_i\) indicates \(B_i[j]\) is not locked, and \(e\) will not visit subsequent layers; (2) \(B_i[j].YES = B_i[j].NO\) indicates a potential replacement, meaning \(e\) will not visit subsequent layers; (3) \(B_i[j].ID = e\) indicates a match with \(e\), and even if the current bucket is locked and cannot be replaced, \(e\) will not visit subsequent layers.





\ppp{Novelty of \anamelong{}.}
The fundamental novelty of \anamelong{} lies in its key idea: identifying keys with significant errors and effectively controlling these errors to completely eliminate outliers. This approach is supported by two innovative techniques we have developed: the Error-Sensible Bucket for error measurement and the Double Exponential Control for error management:

\begin{itemize}
    \item Key Technique I (Error-Sensible Bucket).
    Although the voting technique itself is not novel, tracing back to the classic majority vote \cite{majoritymoore1981fast} algorithm of 1981, our unique contribution lies in demonstrating that the \(NO\) value can effectively limit the extent and impact of collisions. In integrating the Error-Sensible Bucket as the fundamental unit of a sketch, we developed a lock strategy that is directly informed by the size of \(NO\). This approach contrasts with existing works like Majority, MV, Elastic \cite{majoritymoore1981fast, tang2019mvsketch, yang2018elastic}, which primarily concentrate on identifying high-frequency keys but do not fully exploit the capability of \(NO\) in guiding error control.
    \item Key Technique II (Double Exponential Control). To control errors effectively for all keys, it's critical to address the outliers resulting from insertion failures. 
    A crucial strategy is to limit the number of keys advancing to the next layer at each layer. Typically, a layer might halt about half of the keys, which significantly reduces the probability of outlier occurrence, denoted as \(\mathcal{P}\), to \(\frac{1}{2^d}\). Our research indicates that when both the width \(w_i\) and the layer threshold \(\lambda_i\) decrease exponentially, for example, \(\frac{1}{2^{2^{d}}}\) (with \(d=8\), the probability is approximately \(8.6 \times 0.1^{78}\)), the failure probability \(\mathcal{P}\) diminishes at a double exponential rate. This marked decrease in probability effectively reduces the number of keys that can potentially become outliers, thereby eliminating outliers with an extremely high probability.
\end{itemize}

\presub
\subsection{Optimizations and Extensions} \postsub
\label{subsec:alg:extensions}
\label{extension:emergency}

\ppp{Emergency Solution}
For those seeking additional remedies for insertion failures, the following emergency solution can be employed. When an insertion failure occurs, the item \(\langle e, v\rangle\) or a part of it \(\langle e, v'\rangle\) cannot be accommodated in the first \(d\) layers. In such cases, a small hash table or a SpaceSaving structure can be utilized to store these items, ensuring that the portions not inserted are still accurately recorded. This solution is relatively easy to implement on a CPU server. In FPGA or network devices, this additional data structure maintenance can be managed with the help of a CPU-based control plane.
We have implemented these emergency solutions, but chose not to include them in our accuracy evaluation (refer to \cref{sec:experiments}), in order to present the performance of \anamelong{} on its own more clearly.

\ppp{Accuracy Optimization.}
The accuracy of \anamelong{} in practical deployment is a crucial aspect, and our aim is to find optimizations that enhance its accuracy without compromising its excellent theoretical properties.
The first layer of \anamelong{}, which occupies more than 50\% of the entire structure, is its largest. However, this layer can be inefficient when the dataset contains a significant proportion of mice keys (\ie, keys with a small value sum). This is because mice keys sharply increase the NO counters, leading to the locking of most buckets in the first layer, resulting in many buckets being inefficiently used to record these mice keys.
Given that NO counters do not exceed \(\lambda_1\), we propose replacing the first layer with an existing sketch where each counter records up to \(\lambda_1\). This involves substituting each bucket with a counter representing NO, updated with every insertion until reaching \(\lambda_1\). We employed a commonly used CU sketch \cite{estan2002cusketch} for this purpose. In practice, 8-bit counters are adequate for the filter. Compared to a layer consisting of 72-bit error-sensible buckets, this filter can reduce the space requirement of the first layer by nearly 10 times, while introducing only small, manageable errors. However, we cannot replace other layers with this filter because it cannot handle keys with large values.

        \presec
\section{Mathematical Analysis} \postsec
\label{sec:math}

In this section, we provide the key results and key proof steps. 
We have placed the details of non-key steps in our open-source repository, as they are exceedingly complex and we are certain they cannot be fully included in the paper.

\subsection{Key Results}
\label{subsec:math:parametersetting}
\label{key}

We aim to prove the following two key claims.

\ppp{Claim 1:}
The algorithm can achieve the following two goals by using $O\left(\frac{N}{\Lambda}+\ln(\frac{1}{\Delta})\right)$  space:

\begin{align*}
    \Pr\left[\forall \textit{ key } e, \left|\hat{f}(e)-f(e)\right|\leqslant\Lambda\right]\geqslant 1-\Delta
\end{align*}
and 
\begin{align*}
    \forall \textit{ key } e, \Pr\left[\left|\hat{f}(e)-f(e)\right|\leqslant\Lambda\right]\geqslant 1-\Delta
\end{align*}

\ppp{Claim 2:}
The algorithm can achieve the above two goals with $O\left(1+\Delta\ln\ln(\frac{N}{\Lambda})\right)$ amortized time.

\subsection{Key Steps}


Generally, the key steps seek to prove that as \(i\) increases, the number of items entering the \(i\)-th layer during insertion diminishes rapidly. In the \(i\)-th layer, we categorize keys that enter the \(i\)-th layer based on their value size, into elephant keys and mice keys. For elephant keys, we show that their numbers decrease quickly. For mice keys, we show a rapid reduction in their aggregate value. This analysis forms the basis for determining the algorithm's time and space complexity.

Before explaining the proof sketch and key steps, we introduce some basic terms and symbols for clarity. Generally, we assume all values are 1. This means each data item's insertion adds a value of ``1'' to the sketch. The sum \( f(e) \) equals how many times key $e$ is inserted, its frequency. We first prove this for values of 1. Extending the proof to other values is trivial.

When an item is inserted to the \( i \)-th layer and stops the loop, it ``enters'' layers \(1, 2\dots i \) and ``leaves'' layers \(1, 2\dots i-1 \). \( f_i(e) \) represents the times an item with key \( e \) enters layer \( i \). We compare \( f_i(e) \) with \( \frac{\lambda_{i}}{2} \) to categorize keys into two groups at layer \( i \): Mice keys \( \mathcal{S}^{0}_{i} \) and Elephant keys \( \mathcal{S}^{1}_{i} \). We aim to show that the total frequency of mice keys \( F_i \) and the number of distinct elephant keys  \( C_i \) decrease quickly with increasing \( i \). These symbols are detailed in lines 1-5 of Table \ref{table:math:symbols}. For detailed analysis within a bucket, these symbols (lines 2-5) are adapted for the \( j \)-th bucket at layer \( i \) (lines 6-9).

\begin{table}[ht]
\centering
\vspace{-0.05in}
\caption{Common symbols}
\vspace{-0.12in}
\begin{tabular}{l|p{6.7cm}}
\hline
Symbol & Description \\
\hline
(1) $f_i(e)$ & The number of times that key $e$ enters the $i$-th layer. \\
(2) $\mathcal{S}^{0}_{i}$ & $\{e~|~e\in\mathcal{S}_i\land  f_{i}(e)\leqslant \frac{\lambda_{i}}{2}\}$, the set of mice keys. \\
(3) $\mathcal{S}^{1}_{i}$ & $\{e~|~e\in\mathcal{S}_i\land f_{i}(e) > \frac{\lambda_{i}}{2}\}$, the set of elephant keys. \\
(4) $F_i$ & $\sum_{\{e\in\mathcal{S}^{0}_i\}} f_i(e)$, the total frequency of mice keys in $\mathcal{S}^{0}_{i}$. \\
(5) $C_i$ & $|\mathcal{S}^{1}_i|$, the number of elephant keys in $\mathcal{S}^{1}_{i}$. \\
(6) $\mathcal{S}^{0}_{i,j}$ & $\{e~|~e\in\mathcal{S}^{0}_{i}\land h(e)=j\}$, the set of mice keys that are mapped to the $j$-th bucket. \\
(7) $\mathcal{S}^{1}_{i,j}$ & $\{e~|~e\in\mathcal{S}^{1}_{i}\land h(e)=j\}$, the set of elephant keys that are mapped to the $j$-th bucket. \\
(8) $F_{i,j}$ & $\sum_{\{e\in\mathcal{S}^{0}_{i,j}\}} f_i(e)$, the total frequency of mice keys in $\mathcal{S}^{0}_{i,j}$. \\
(9) $C_{i,j}$ & $|\mathcal{S}^{1}_{i,j}|$, the number of elephant keys in $\mathcal{S}^{1}_{i,j}$. \\
(10) $\mathcal{P}_{i,k}$ & $\{e_1,\cdots,e_{k}\}$, a subset of $\mathcal{S}_i$ composed of the first $k$ keys. \\
(11) $f^{P}_{i,k}$ & $\sum_{\{e\in\mathcal{P}_{i,k-1}\cap\mathcal{S}^{0}_{i,h(e_k)}\}} f_i(e)$, the total frequency of mice keys with a smaller index that conflicts with key $e_k$. \\
(12) $c^{P}_{i,k}$ & $\left|\{e~|~e\in\mathcal{P}_{i,k-1}\cap\mathcal{S}^{1}_{i,h(e_k)}\}\right|$, the number of elephant keys with a smaller index that conflicts with key $e_k$. \\
\hline
\end{tabular}
\label{table:math:symbols}
\vspace{-0.15in}
\end{table}

\ppp{Proof sketch:}
The proof consists of the following four steps.
The first three steps focus on a single layer (the $i$-th layer), and analyze the relationship between the $i$-th layer and the $(i+1)$-th layer.
The fourth step traverses all layers to draw a final conclusion.
\begin{itemize}[leftmargin=*]
     \item  \textbf{Step 1 (Bound mice and elephant keys leaving the $i$-th layer with Xi and Yi, respectively)}.
     Our analysis must be applicable regardless of the order in which any item is inserted into the sketch. We start by analyzing, among the items entering the $i$-th layer, how many will proceed to the \((i+1)\)-th layer, thereby leaving the $i$-th layer. We construct two time-order-independent random variables \(X_i\) and \(Y_i\) to bound mice keys and elephant keys, respectively: \(X_i\) bounds the total frequency of the mice keys leaving the \(i\)-th layer, and \(Y_i\) bounds the number of distinct elephant keys leaving the \(i\)-th layer (Theorem \ref{Theorem:x:y}).
      
     \item  \textbf{Step 2 (Double exponential decrease of \(X_i\) and \(Y_i\))}: We prove that if the number of mice keys \( F_i \) and elephant keys \( C_i \) in the \(i\)-th layer decrease double exponentially, then the quantity of keys leaving the \(i\)-th layer, \ie, \(X_i\) and \(Y_i\), will also decrease double exponentially (Theorem \ref{app:theo:exp_1}).
    
     \item  \textbf{Step 3 (Double exponential decrease of \(F_{i+1}\) and \(C_{i+1}\))}: Although the quantities of \(X_i\) and \(Y_i\) leaving the \(i\)-th layer are within limits, it does not directly imply that the number of mice keys \( F_{i+1} \) and elephant keys \( C_{i+1} \) in the \((i+1)\)-th layer are few. This is because the criteria for categorizing an elephant key differ across layers, and a mice key from the \(i\)-th layer may become an elephant key upon entering the \((i+1)\)-th layer. Fortunately, by using a Concentration inequality, we prove that this situation is controllable. That is, if \( F_i \) and \( C_i \) decrease double exponentially, \( F_{i+1} \) and \( C_{i+1} \) in the next layer will also decrease double exponentially (Theorem \ref{theo:gamma}).
    
     \item  \textbf{Step 4 (Combine all layers)}:
    Based on step 3, by using Boole's inequality, we combine the results from each layer and prove that there is a high probability (\(1-\Delta\)) that the final conclusion holds (Theorem \ref{theo:main}).

\end{itemize}

\ppp{The results of step 1.  }


\pretheo\begin{theorem}
\label{Theorem:x:y}    
    Let
    \begin{align*}
        X_{i,k}=
            \begin{cases}
              0 & C_{i, h(e_k)}=0 \land f^{P}_{i,k}\leqslant \frac{\lambda_i}{2}
            \\
              f_i(e_k) &  C_{i, h(e_k)}=0\land f^{P}_{i,k} > \frac{\lambda_i}{2}
            \\
              f_i(e_k) & C_{i, h(e_k)}>0
            \end{cases},
        &&
        X_i=\sum_{\{e_k\in\mathcal{S}^{0}_{i}\}} X_{i,k}.
    \end{align*}
    The total frequency of the mice keys leaving the $i$-th layer does not exceed $X_i$, \ie,
    \begin{align*}
        F_{i+1}
        \leqslant
        \sum_{\{e\in\mathcal{S}^{0}_i\cap\mathcal{S}_{i+1}\}} f_{i+1}(e) \leqslant X_{i}.
    \end{align*}

   Let 
    \begin{align*}
        &Y_{i,k}=
            \begin{cases}
              0 &  c^{P}_{i,k}=0\land F_{i,h(e_k)}\leqslant \lambda_i,
            \\
              2 & c^{P}_{i,k}=0\land F_{i,h(e_k)}> \lambda_i
            \\
              2 & c^{P}_{i,k}>0.
            \end{cases},
        &&
        Y_{i}=\sum_{e_k\in\mathcal{S}^{1}_{i}} Y_{i,k}.
    \end{align*}
    The number of distinct elephant keys leaving the $i$-th layer does not exceed $Y_i$, \ie,
    \begin{align*}
        \left|\mathcal{S}^{1}_{i}\cap\mathcal{S}^{1}_{i+1}\right|\leqslant Y_{i}.
    \end{align*}
\end{theorem}\posttheo

\ppp{The results of step 2.}

\pretheo\begin{theorem}
\label{app:theo:exp_1}
Let
$W=\frac{4N(R_wR_\lambda)^{6}}{\Lambda(R_w-1)(R_\lambda-1)}$,
$\alpha_i= \frac{\Vert F\Vert_1}{(R_wR_\lambda)^{i-1}}$,
$\beta_i= \frac{\alpha_i}{\frac{\lambda_i}{2}}$,
$\gamma_i=(R_wR_\lambda)^{(2^{i-1}-1)}$, and
$p_i = (R_wR_\lambda)^{-(2^{i-1}+4)}$.
Under the conditions of
    $F_i\leqslant\frac{\alpha_i}{\gamma_i}$ and
    $C_i\leqslant\frac{\beta_i}{\gamma_i}$,
we have
\begin{align*}
    \Pr\left(X_{i}>(1+\Delta)\frac{p_i\alpha_i}{\gamma_i}\right)
    \leqslant
    \exp\left(-(\Delta-(e-2))\frac{2p_i\alpha_i}{\lambda_i\gamma_i}\right).
\end{align*}
and
\begin{align*}
    \Pr\left(Y_{i}>(1+\Delta)\frac{3}{2}\frac{p_i\beta_i}{\gamma_i}\right)
    \leqslant
    \exp\left(-(\Delta-(e-2))\frac{3p_i\beta_i}{4\gamma_i}\right).
\end{align*}

\end{theorem}\posttheo

\ppp{The results of step 3.}



\pretheo\begin{theorem}
\label{theo:gamma}
Let
$R_wR_\lambda\geqslant2$,
$W=\frac{4N(R_wR_\lambda)^{6}}{\Lambda(R_w-1)(R_\lambda-1)}$,
$\lambda_i = \frac{\Lambda (R_\lambda-1)}{R_\lambda^i}$,
$\alpha_i= \frac{N}{(R_wR_\lambda)^{i-1}}$,
$\beta_i= \frac{\alpha_i}{\frac{\lambda_i}{2}}$,
$\gamma_i=(R_wR_\lambda)^{(2^{i-1}-1)}$, and
$p_i = (R_wR_\lambda)^{-(2^{i-1}+4)}$.
We have
\begin{align*}
&    
    \Pr\left(F_{i+1}>\frac{\alpha_{i+1}}{\gamma_{i+1}}
    ~|~
    F_{i}\leqslant\frac{\alpha_{i}}{\gamma_{i}} \land C_{i}\leqslant\frac{\beta_{i}}{\gamma_{i}}\right)
    \leqslant
    \exp\left(-(9-e)\frac{2p_i\alpha_i}{\lambda_i\gamma_i}\right).
    \\
&
    \Pr\left(C_{i+1}>\frac{\beta_{i+1}}{\gamma_{i+1}}
    ~|~
    F_{i}\leqslant\frac{\alpha_{i}}{\gamma_{i}} \land
    C_{i}\leqslant\frac{\beta_{i}}{\gamma_{i}}\right)
    \\
    \leqslant&
    \exp\left(-(5-e)\frac{2p_i\alpha_i}{\lambda_i\gamma_i}\right)
    +
    \exp\left(-(\frac{11}{3}-e)\frac{3p_i\beta_i}{4\gamma_i}\right).
\end{align*}
\end{theorem}\posttheo


\ppp{The results of step 4.}

\pretheo\begin{theorem}
\label{theo:main}
Let
$R_wR_\lambda\geqslant 2$,
$W=\frac{4N(R_wR_\lambda)^{6}}{\Lambda(R_w-1)(R_\lambda-1)}$,
$\lambda_i = \frac{\Lambda (R_\lambda-1)}{R_\lambda^i}$,
$\alpha_i= \frac{N}{(R_wR_\lambda)^{i-1}}$,
$\beta_i= \frac{\alpha_i}{\frac{\lambda_i}{2}}$,
$\gamma_i=(R_wR_\lambda)^{(2^{i-1}-1)}$, and
$p_i = (R_wR_\lambda)^{-(2^{i-1}+4)}$.
For given $\Lambda$ and $\Delta<\frac{1}{4}$, 
let $d$ be the root of the following equation
\begin{align*}
    \frac{R_\lambda^{d}}{(R_wR_\lambda)^{(2^{d}+d)}}=\Delta_1\frac{\Lambda}{N}\ln(\frac{1}{\Delta}).
\end{align*}
And use an SpaceSaving of size $\Delta_2\ln(\frac{1}{\Delta})$ (as the $(d+1)$-layer), then 
\begin{align*}
    \Pr\left(\forall \textit{ key } e, \left|\hat{f}(e)-f(e)\right|\leqslant\Lambda\right)\geqslant 1-\Delta,
\end{align*}
where
\begin{align*}
    \Delta_1=2R_w^2R_\lambda^2(R_\lambda-1),
    &&
    \Delta_2=3\left(\frac{R_wR_\lambda^2}{R_\lambda-1}\right)
    \Delta_1=6R_w^3R_\lambda^4.
\end{align*}
\end{theorem}\posttheo


\ppp{Complexity of \aname{}.}

\pretheo\begin{theorem}
Using the same settings as Theorem \ref{theo:main}, the space complexity of the algorithm is $O(\frac{N}{\Lambda}+\ln(\frac{1}{\Delta}))$, and the time complexity of the algorithm is amortized $O(1+\Delta\ln\ln(\frac{N}{\Lambda}))$.
\end{theorem}\posttheo

\presec
\section{Implementations} \postsec
\label{sec:Implementations}

We have implemented \anamelong{} on three platforms: CPU server, FPGA, and Programmable Switch. Given the challenging nature of implementations on the latter two platforms, due to various hardware constraints, we provide a brief introduction here. Our source code is available on GitHub \cite{opensource}.

\presub
\subsection{FPGA} \postsub
\label{sec:FPGA:deploy}
We implement the \aname{} on an FPGA network experimental platform (Virtex-7 VC709). The FPGA integrated with the platform is xc7vx690tffg1761-2 with 433200 Slice LUTs, 866400 Slice Register, and 1470 Block RAM Tile. The implementation mainly consists of three hardware modules: calculating hash values (hash), Error-Sensible Buckets (ESbucket), and a stack for emergency solution (Emergency). \anamelong{} is fully pipelined, which can input one key in every clock, and complete the insertion after 41 clocks. According to the synthesis report (see Table 1), the clock frequency of our implementation in FPGA is 340 MHz, meaning the throughput of the system can be 340 million insertions per second.

\begin{table} [h]
\renewcommand\arraystretch{1.05}
\vspace{-0.12in}
    \caption{FPGA Implementation Results.}
    \label{tb:FPGA}
    \vspace{-0.15in}
    \begin{tabular}{m{0.2\columnwidth}|m{0.12\columnwidth}<{\raggedleft}|m{0.15\columnwidth}<{\raggedleft}|m{0.12\columnwidth}<{\raggedleft}|m{0.2\columnwidth}<{\raggedleft}}
\bottomrule[1pt]
    \textbf{Module} & \centering \makecell{\textbf{CLB}\\\textbf{LUTS}} & \makecell{\textbf{CLB}\\\textbf{Register}} &\makecell{\textbf{Block}\\\textbf{RAM}} &\makecell{\textbf{Frequency}\\\textbf{(MHz)}}\\
    \hline
    Hash & 85 & 130 & 0 & 339\\
    ESbucket &  2521 &  2592 & 258 & 339\\
    Emergency &  48 &  112 & 1 & 339\\ \hline
     \specialrule{0em}{1pt}{1pt} \hline
    Total &  2654 &  2834 & 259 & 339\\
    Usage &  0.61\% &  0.33\% & 17.62\% & \\
\toprule[1pt]
    \end{tabular}

\vspace{-0.12in}
\end{table}

\subsection{Programmable Switch}

To implement \aname{} on programmable switches (\eg, Tofino), we need to solve the following three challenges.

\ppp{Challenge I: Circular Dependency.}
Programmable switches limit SALU access to a pair of 32-bit data per stage, but each \anamelong{} bucket contains three fields (\texttt{ID}, \texttt{YES}, \texttt{NO}), creating dependencies that exceed this limit. To resolve this, we simplify the dependencies by using the difference between \texttt{YES} and \texttt{NO} (\texttt{DIFF}) for replacement decisions. This adjustment allows us to align \texttt{DIFF} and \texttt{ID} in the first stage and \texttt{NO} in the second stage, breaking the dependency cycle.

\ppp{Challenge II: Backward Modification.}
When \texttt{NO} surpasses a threshold, the bucket must be locked, preventing updates to \texttt{ID}. However, due to pipeline constraints, a packet can’t modify the \texttt{LOCKED} flag within its lifecycle. Our solution involves recirculating the packet that first exceeds the threshold, allowing it to re-enter the pipeline and update the flag.

\ppp{Challenge III: Three Branches Update and Output Limitation.}
Weighted updates to \texttt{DIFF} could result in three different values, but switches can only support two variations. To accommodate this limitation and the 32-bit output constraint per stage, we simplify the update process. When not matching \texttt{ID}, \texttt{DIFF} is updated using saturated subtraction. In replacement scenarios, \texttt{DIFF} is reduced to zero, and \texttt{ID} is replaced upon the arrival of the next packet identifying \texttt{DIFF} as zero.

\begin{table}[H]
\vspace{-0.12in}
\caption{H/W Resources Used by \aname{}. }
\label{table:resource} 
\vspace{-0.3cm}
\begin{tabular}{m{0.35\columnwidth}|m{0.25\columnwidth}<{\raggedleft}|m{0.27\columnwidth}<{\raggedleft}}
\bottomrule[1pt]
\textbf{Resource}   & \textbf{Usage} & \textbf{Percentage} \\ \hline
Hash Bits  & 541 & 10.84\% \\
SRAM  & 138 & 14.37\% \\
Map RAM  & 119 & 20.66\% \\
TCAM  & 0 & 0\% \\
Stateful ALU  & 12 & 25.00\% \\
VLIW Instr  & 23 & 5.99\% \\
Match Xbar  & 109 & 7.10\% \\
 \toprule[1pt]
\end{tabular}
\vspace{-0.05in}
\end{table}

\ppp{Hardware resource utilization:}
%
After solving the above three challenges, we have fully implemented \aname{} on Edgecore Wedge 100BF-32X switch (with Tofino ASIC).
Table \ref{table:resource} lists the utilization of various hardware resources on the switch.
The two most used resources of \aname{} are Map RAM and Stateful ALU, which are used 20.66\% and 25\% of the total quota, respectively.
These two resources are mainly used by the multi-level bucket arrays in \aname{}.
For other kinds of sources, \aname{} uses up to 14.37\% of the total quota.



	\presec
\section{Experiment Results} \postsec
\label{sec:experiments}


In this section, we present the experiment results for \anamelong{}. We begin by the setup of the experiments (\cref{subsec:exp:setup}). Following this, we perform a comparative analysis of \anamelong{} against existing solutions in terms of accuracy (\cref{subsec:exp:accuracycompare}) and speed (\cref{subsec:exp:Speedcompare}). 
Finally, we evaluate \anamelong{} in detail, including the impact of its parameters on performance (\cref{subsec:exp:Parameters}), its capability in error sensing and control, and industrial deployment performance (\cref{subsec:exp:Observations}).
The source code is available on GitHub \cite{opensource} under anonymous release.



\presub
\subsection{Experiment Setup} \postsub
\label{subsec:exp:setup}



\subsubsection{Implementation}
Our experiments are mostly based on C++ implementations of \anamelong{} and related algorithms.
Here we use fast 32-bit Murmur Hashing \cite{murmur}, and different hash functions that affect accuracy little. Each bucket of \aname{} consists of a 32-bit $YES$ counter, a 16-bit $NO$ counter, and a 32-bit $ID$ field. Mice filter occupies 20\% of total memory, and bucket size of it is fixed to 2 bits unless otherwise noted. 
According to the study in \cref{subsec:exp:Parameters}, we set $R_w$ to 2 and $R_\lambda$ to 2.5 by default. The memory size is 1MB and the user-defined threshold $\Lambda$ is 25 by default.
All the experiments are conducted on a server with 18-core CPU (36 threads, Intel CPU i9-10980XE @3.00 GHz), which has 128GB memory. Programs are compiled with O2 optimization.


\subsubsection{Datasets}
We use four large-scale real-world streaming datasets and one synthetic dataset, with the first dataset being the default. 
\begin{itemize}[leftmargin=0.15in]
    \item \textbf{IP Trace (Default): } An anonymized dataset collected from \cite{caida}, comprised of IP packets. We use the source and destination IP addresses as the key. The first 10M packets of the whole trace are used to conduct experiments, including about 0.4M distinct keys.
    \item \textbf{ Web Stream: } A dataset built from a spidered collection of web HTML documents \cite{webpage}. The first 10M items of the entire trace are used to conduct experiments, including about 0.3M distinct keys.
    \item \textbf{University Data Center: } An anonymized packet trace from university data center \cite{benson2010network}. We fetch 10M packets of the dataset, containing about 1M distinct keys.
    \item \textbf{Hadoop Stream: } A dataset built from real-world traffic distribution of HADOOP. The first 10M packets of the whole trace are used to conduct experiments, including about 20K distinct keys.
    \item \textbf{Synthetic Datasets:} We generate \cite{webpoly} several synthetic datasets according to a Zipf distribution with different skewness for experiments, each of them consists of 32M items.
\end{itemize}
By default, the value is set to 1 to allow for comparison with existing methods, unless the unit of the value is explicitly mentioned.



\subsubsection{Evaluation Metrics}
\label{subsec:exp:metric}
We evaluate the performance of \aname{} and its competitors using the following four metrics. Given our objective to control all errors below the user-defined threshold, our accuracy evaluation focuses more on the first metric, \# Outliers, rather than metrics like AAE.

\begin{itemize}[leftmargin=0.15in]
    \item \ppp{The Number of Outliers (\# Outliers):} 
    The number of keys whose absolute error of estimation is greater than the user-defined threshold $\Lambda$.
    \item \ppp{Average Absolute Error (AAE):} $\frac{1}{|U|} \sum\limits_{e_i \in U}|f(e_i)-\hat{f}(e_i)| $,
where $U$ is the set of keys, $f(e_i)$ is the true value sun of key $e_i$, and $\hat{f}(e_i)$ is the estimation. 
    \item \ppp{Average Relative Error (ARE):} $\frac{1}{|U|} \sum\limits_{e_i \in U}\frac{|f(e_i)-\hat{f}(e_i)|}{f(e_i)} $.
    \item \ppp{Throughput:} $\frac{N}{T}$, where $N$ is the number of operations and $T$ is the elapsed time. Throughput describes the processing speed of an algorithm, and we use Million of packets per second (Mpps) to measure the throughput.
\end{itemize}

\subsubsection{Implementation of Competitor}
We conduct experiments to compare the performance of \aname{} ("Ours" in figures) with seven competitors, including CM \cite{cmsketch}, CU \cite{estan2002cusketch}, SS \cite{spacesaving}, Elastic \cite{yang2018elastic}, Coco \cite{zhang2021cocosketch}, HashPipe \cite{hashpipe}, and PRECISION\cite{ben2018efficient}. 
For CM and CU, we provide fast (CM\_fast/CU\_fast) and accurate (CM\_acc/CU\_acc) two versions, implementing 3 and 16 arrays respectively.
For Elastic, its light/heavy memory ratio is 3 as recommended \cite{yang2018elastic}.
For Coco, we set the number of arrays $d$ to 2 as recommended \cite{zhang2021cocosketch}.
For HashPipe, we set the number of pipeline stages $d$ to 6 as recommended \cite{hashpipe}.
And for PRECISION, we set the number of pipeline stages $d$ to 3 for best performance \cite{ben2018efficient}.








\presub
\subsection{Accuracy Comparison} \postsub
\label{subsec:exp:accuracycompare}

\aname{} controls error efficiently as our expectation and achieves the best accuracy compared with competitors. 
In evaluating accuracy, we considered three aspects: the number of outliers in all keys, the number of outliers in frequent keys, and average estimation error of values.

\presubsub
\subsubsection{Number of Outliers in All Keys.}
\label{subsubsec:exp:Outliersall}

Under various \(\Lambda\) values and across different datasets, we consistently achieve zero outliers with more than 2 times memory saving.

\begin{figure}[htbp]
\prefig
    \centering
 	    \begin{subfigure}{0.49\linewidth}
        \centering
		\includegraphics[width=\textwidth, ]{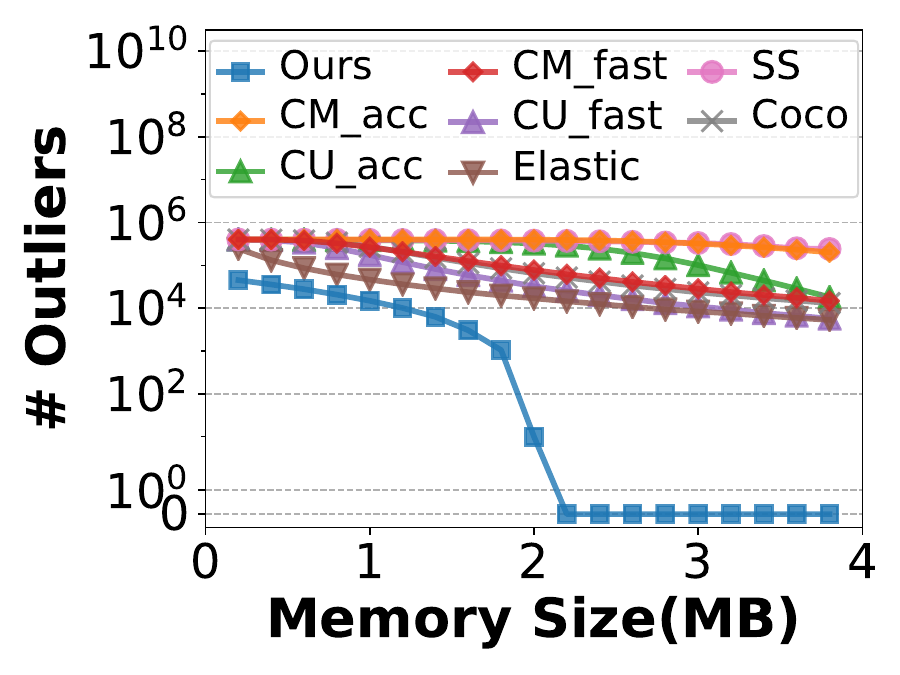}
        \presubfigcaption
        \caption{$\Lambda$=5}
		\label{subfig:app:range5:caida}
        \end{subfigure}
    \hfill
 	    \begin{subfigure}{0.49\linewidth}
        \centering
		\includegraphics[width=\textwidth, ]{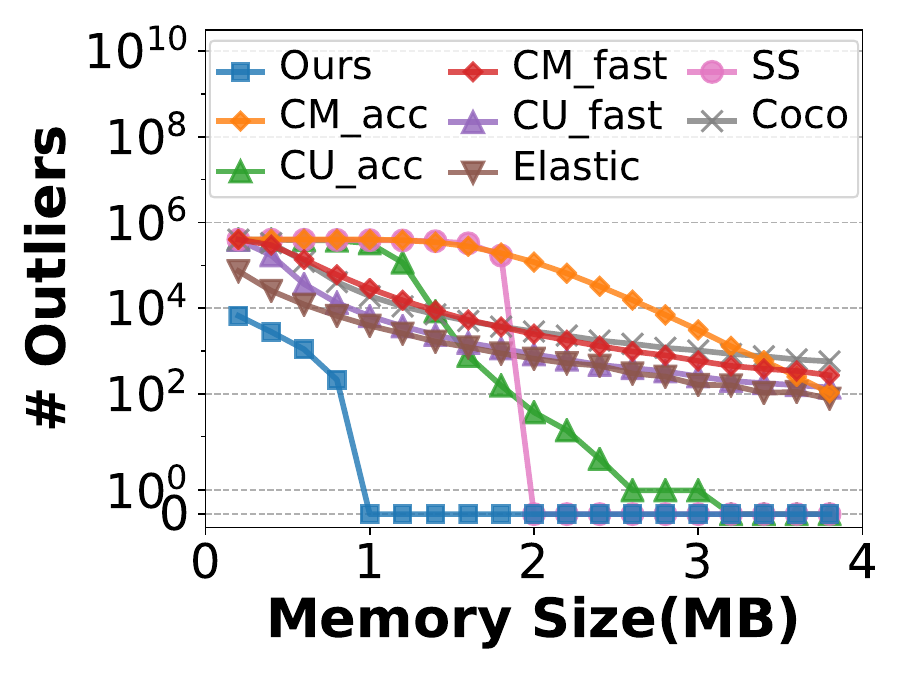}
        \presubfigcaption
        \caption{$\Lambda$=25}
		\label{subfig:app:range25:caida}
        \end{subfigure}
        
    \prefigcaption
    \caption{\# Outliers in Different $\Lambda$.}
	\label{fig:app:range_eps}
\postfig
\end{figure}

\ppp{Impact of Threshold $\Lambda$ (Figure \ref{subfig:app:range5:caida}, \ref{subfig:app:range25:caida}):}
We vary $\Lambda$ and count the number of outliers.
As the figures show, \aname{} takes the lead position regardless of $\Lambda$.
When $\Lambda$=$25$, \aname{} achieves zero outlier within 1MB memory, while the others still report over 5000 outliers. 

\begin{figure}[htbp]
\prefig
    \centering
	\includegraphics[width=0.85\linewidth, ]{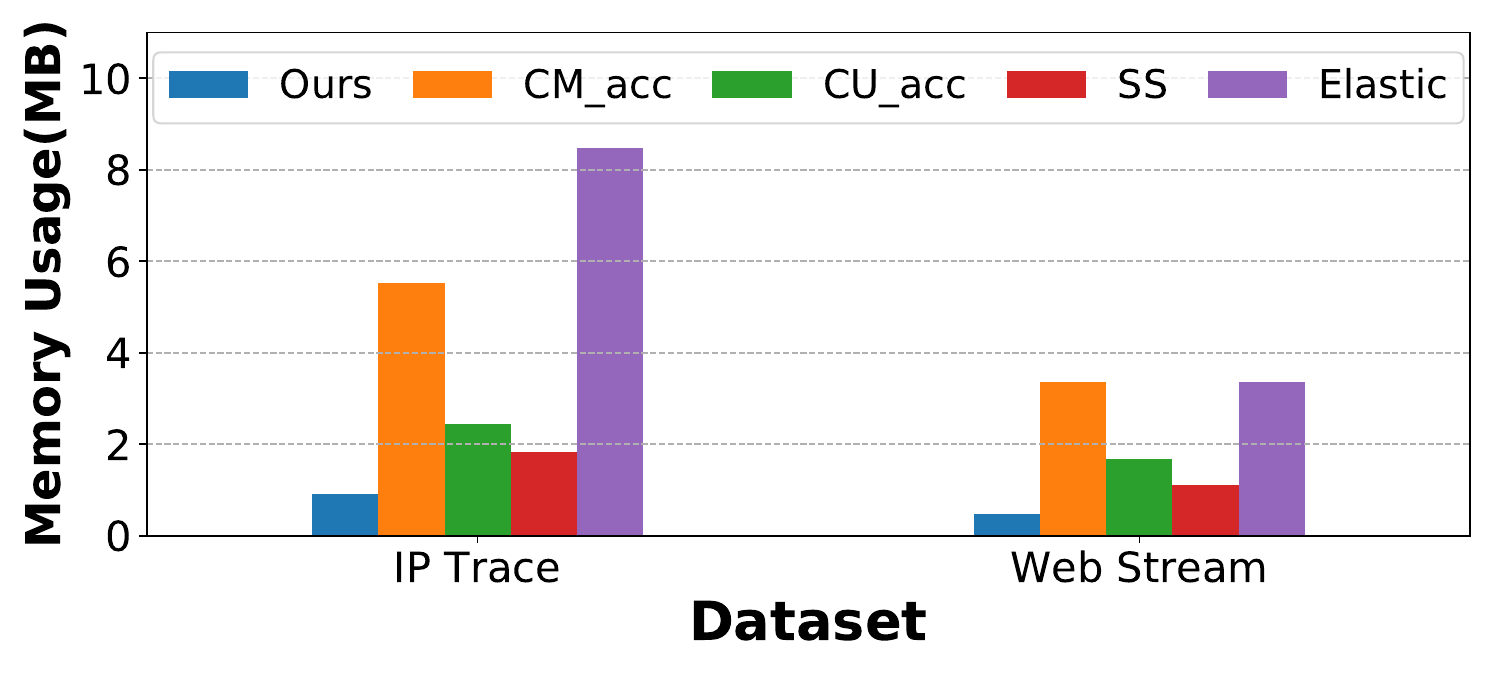}
    \prefigcaption
    \caption{Memory Consumption under Zero Outlier.}
	\label{fig:app:outlier_minmem}
\postfig
\end{figure}

\ppp{Zero-Outlier Memory Consumption (Figure \ref{fig:app:outlier_minmem}):} We further explore the precise minimum memory consumption to achieve zero outlier for all algorithms, $\Lambda$ is fixed to 25 and experiments are conducted on different datasets. 
For the IP Trace dataset, memory consumption of \aname{} is 0.91MB, about 6.07, 2.69, 2.01, 9.32 times less than CM (accurate), CU (accurate), Space-Saving, and Elastic respectively.
CM (fast), CU (fast) and Coco cannot achieve zero outlier within 10MB memory.
Besides, CM, CU and Elastic usually require more memory than the minimum value, otherwise they cannot achieve zero outlier stably.

\begin{figure}[htbp]
    \centering
 	    \begin{subfigure}{0.49\linewidth}
        \centering
		\includegraphics[width=\textwidth, ]{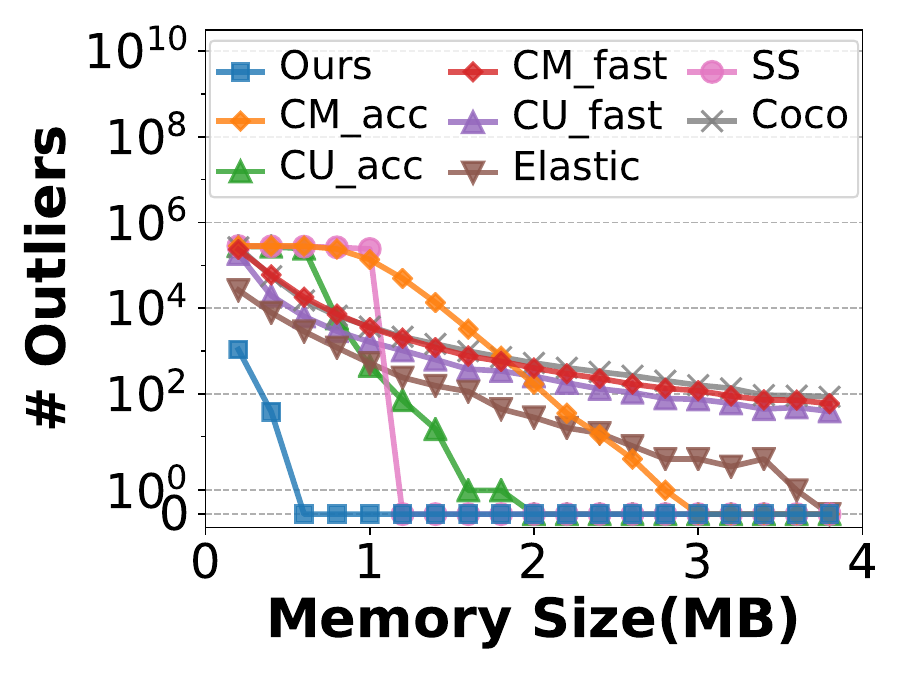}
        \presubfigcaption
        \caption{Web Stream}
		\label{subfig:app:range25:webpage}
        \end{subfigure}
    \hfill
 	    \begin{subfigure}{0.49\linewidth}
        \centering
		\includegraphics[width=\textwidth, ]{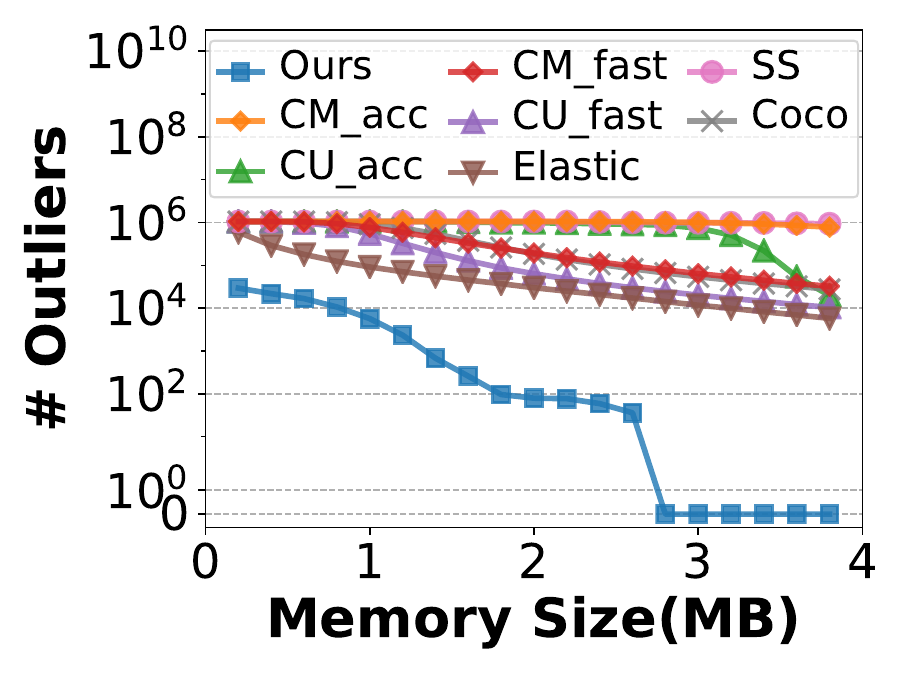}
        \presubfigcaption
        \caption{University Data Center}
		\label{subfig:app:range25:kosarak}
        \end{subfigure}
        
    \centering
 	    \begin{subfigure}{0.49\linewidth}
        \centering
		\includegraphics[width=\textwidth, ]{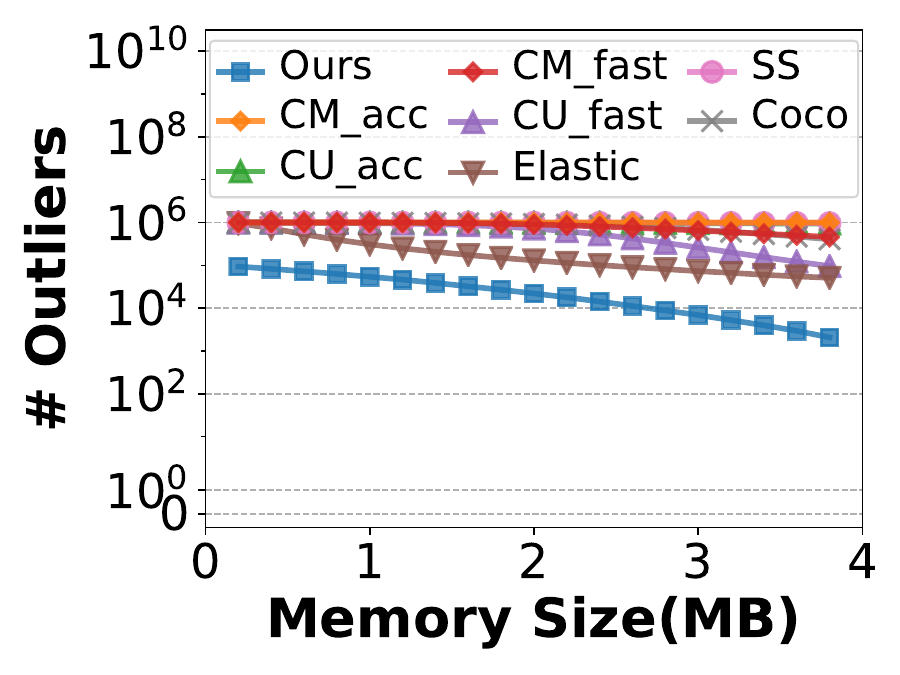}
        \presubfigcaption
        \caption{Synthetic (Skewness=0.3)}
		\label{subfig:app:range25:syn003}
        \end{subfigure}
    \hfill
 	    \begin{subfigure}{0.49\linewidth}
        \centering
		\includegraphics[width=\textwidth, ]{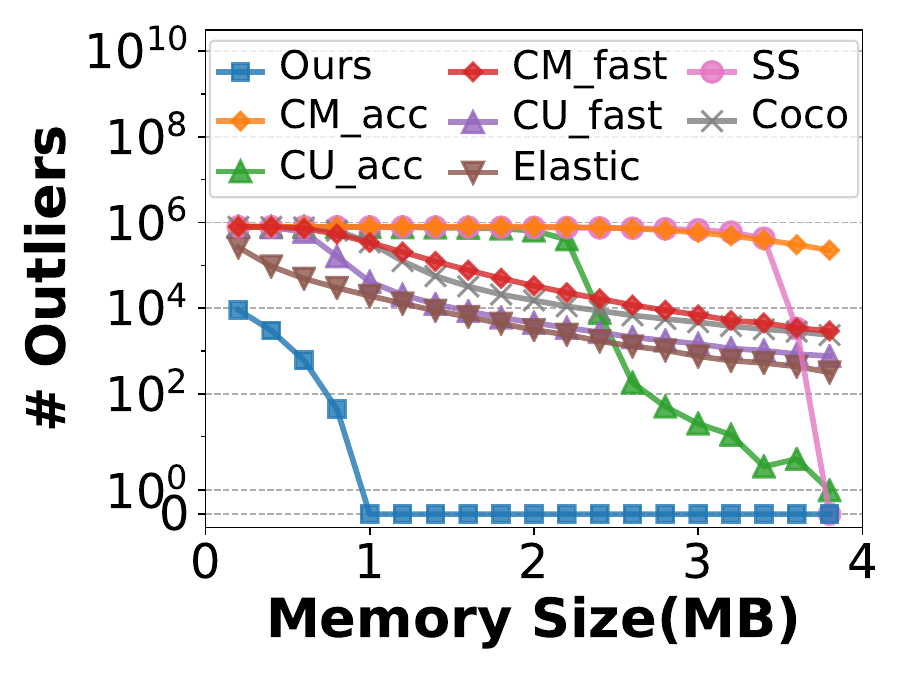}
        \presubfigcaption
        \caption{Synthetic (Skewness=3.0)}
		\label{subfig:app:range25:syn030}
        \end{subfigure}
        
    \prefigcaption
    \caption{\# Outliers on Different Datasets.}
	\label{fig:app:range_dataset}
\postfig
\end{figure}

\ppp{Impact of Dataset (Figure \ref{subfig:app:range25:webpage},\ref{subfig:app:range25:kosarak},\ref{subfig:app:range25:syn003},\ref{subfig:app:range25:syn030}):} We fix $\Lambda$ to 25 and change the dataset. The figures illustrate that \aname{} has the least memory requirement regardless of the dataset. 
For synthetic dataset with skewness=0.3, no algorithm achieves zero outlier within 4MB memory, while the number of outliers of \aname{} is over 50 times less than others.

\presubsub
\subsubsection{Number of Outliers in Frequent Keys.}
\label{subsubsec:exp:Outliersall}

\begin{figure}[htbp]
\prefig
    \centering
 	    \begin{subfigure}{0.47\linewidth}
        \centering
		\includegraphics[width=\textwidth, ]{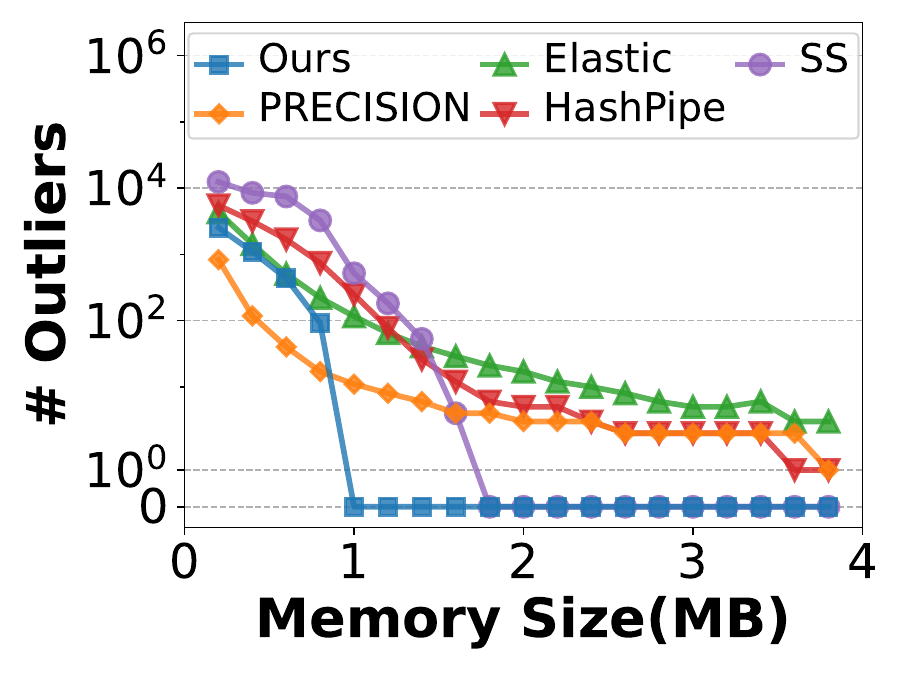}
        \presubfigcaption
        \caption{$T = 100$}
		\label{subfig:app:hh100:caida}
        \end{subfigure}
    \hfill
 	    \begin{subfigure}{0.47\linewidth}
        \centering
		\includegraphics[width=\textwidth, ]{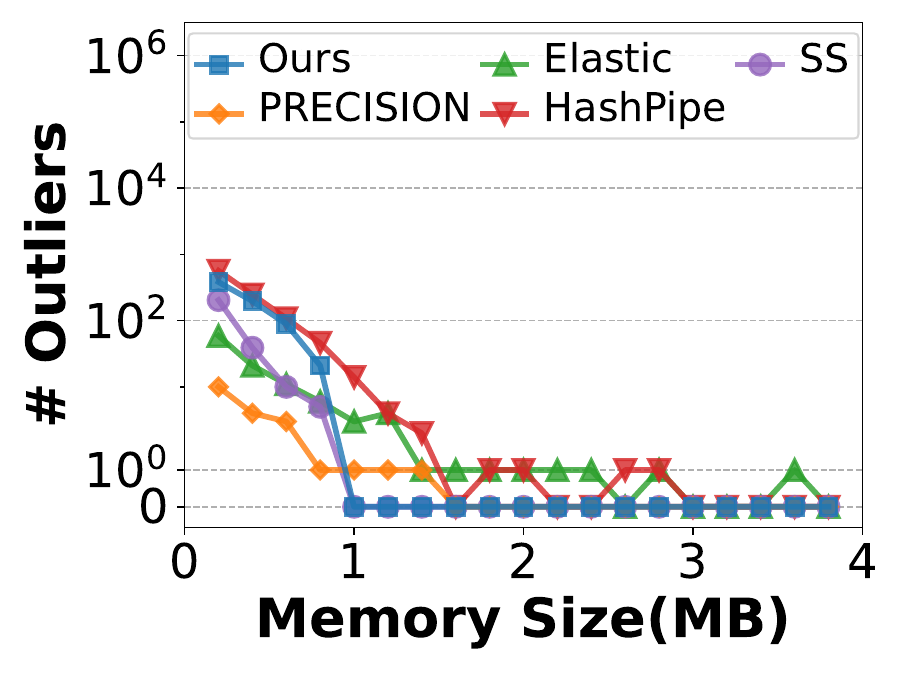}
        \presubfigcaption
        \caption{$T = 1000$}
		\label{subfig:app:hh1000:caida}
        \end{subfigure}
        
    \prefigcaption
    \caption{Number of Outliers in Elephant Keys.}
	\label{fig:app:hh}
\postfig
\end{figure}

We define keys with a value sum greater than \(T\) as frequent keys and count the outliers among them, as frequent keys often draw more attention. 
We varied the memory usage from 200KB to 4MB, using the number of outliers to measure the accuracy of the aforementioned algorithms. The user-defined \(\Lambda\) is set at 25. To clearly demonstrate performance under extreme confidence levels, we changed the hash seed, conducted 100 repeated experiments for each specific setting, and presented the worst-case scenario.

Figure \ref{fig:app:hh} shows that \anamelong{} requires the least memory to achieve zero outliers with stable performance. For \(T = 100\), there are a total of 12,718 frequent keys. Space-Saving needs about 1.8 times more memory than \anamelong{}, and other solutions cannot eliminate outliers within 4MB. For \(T = 1000\), with 1,625 frequent keys, \anamelong{} achieves performance comparable to Space-Saving. However, it's noteworthy that Space-Saving cannot be implemented on hardware platforms, including programmable switches and FPGAs.

\subsubsection{Average Error.}

Average error measures the average difference between estimated and actual values. \aname{} is comparable to the best solutions in this regard. However, optimizing average error is not our primary goal, because its correlation with confidence is relatively low.

\begin{figure}[htbp]
        \centering
             \begin{subfigure}{0.47\linewidth}
            \centering
            \includegraphics[width=\textwidth, ]{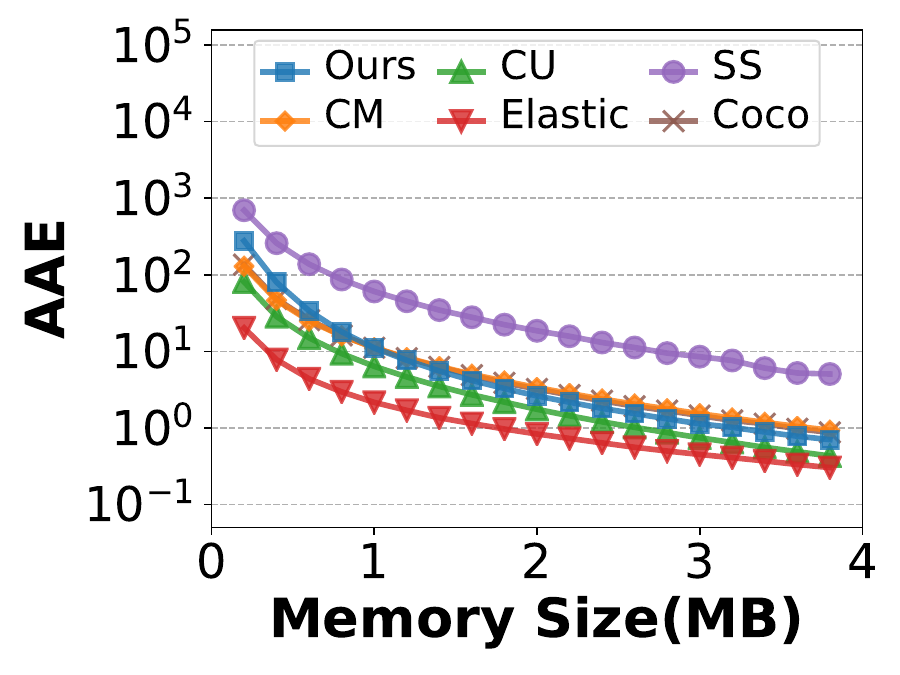}
            \presubfigcaption
            \caption{IP Trace}
            \label{subfig:app:fe_aae:caida}
            \end{subfigure}
        \hfill
             \begin{subfigure}{0.47\linewidth}
            \centering
            \includegraphics[width=\textwidth, ]{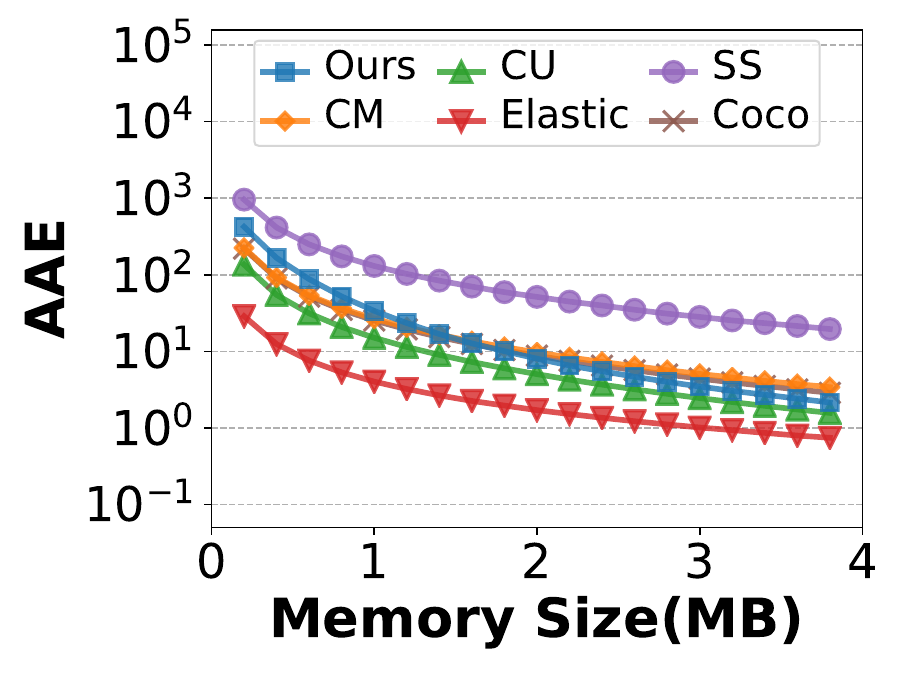}
            \presubfigcaption
            \caption{Synthetic (Skewness=3.0)}
            \label{subfig:app:fe_aae:syn030}
            \end{subfigure}
            
        \prefigcaption
        \caption{AAE on Different Datasets.}
        \label{fig:exp_app:aae}
    \postfig
    \end{figure}
    
    \begin{figure}[htbp]
    \prefig
        \centering
             \begin{subfigure}{0.47\linewidth}
            \centering
            \includegraphics[width=\textwidth, ]{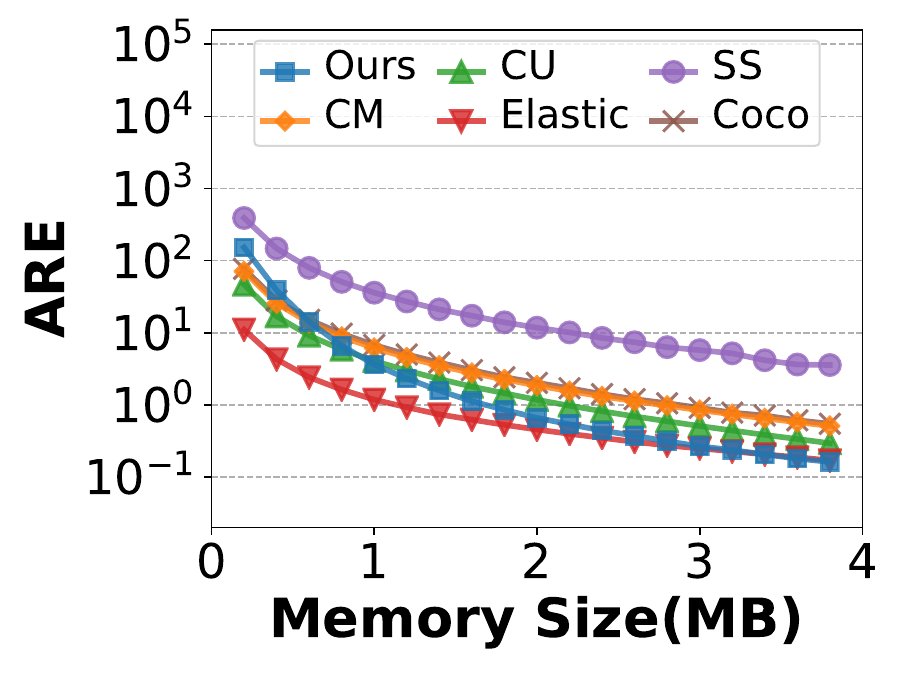}
            \presubfigcaption
            \caption{IP Trace}
            \label{subfig:app:fe_are:caida}
            \end{subfigure}
        \hfill
             \begin{subfigure}{0.47\linewidth}
            \centering
            \includegraphics[width=\textwidth, ]{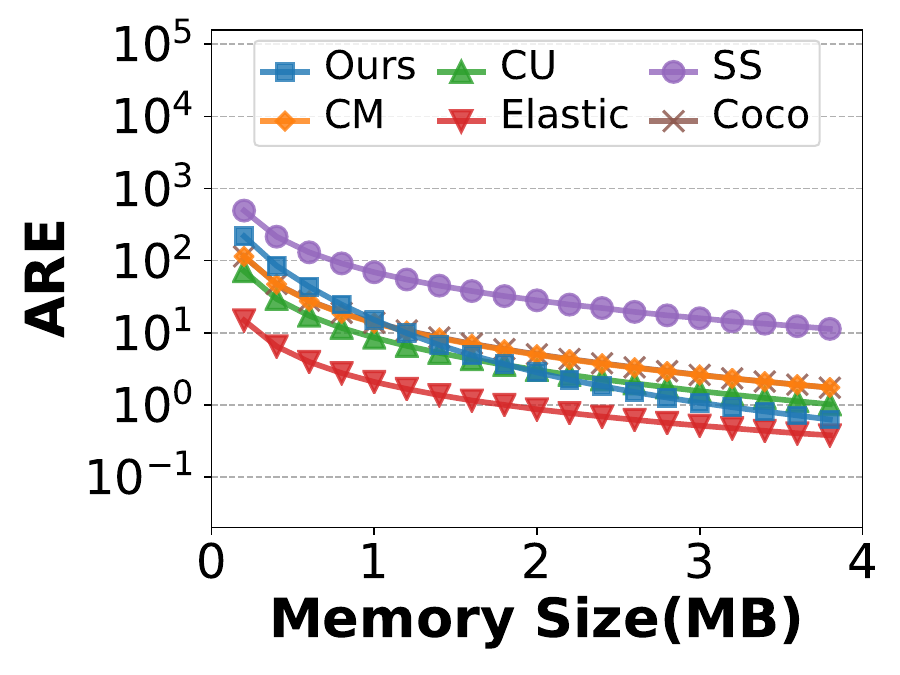}
            \presubfigcaption
            \caption{Synthetic (Skewness=3.0)}
            \label{subfig:app:fe_are:syn030}
            \end{subfigure}
        \prefigcaption
        \caption{ARE on Different Datasets.}
        \label{fig:exp_app:are}
    \postfig
\end{figure}



\ppp{AAE \emph{vs.} Memory Size (Figure \ref{subfig:app:fe_aae:caida}, \ref{subfig:app:fe_aae:syn030}):}
It is shown that when memory size is up to 4MB, \aname{} has a comparable AAE with Elastic and CU in two datasets, is about $1.59 \sim 2.01$ times lower than CM, $1.34 \sim 1.69$ times lower than Coco, and $9.10 \sim 11.48$ times lower than Space-Saving.

\ppp{ARE \emph{vs.} Memory Size (Figure \ref{subfig:app:fe_are:caida}, \ref{subfig:app:fe_are:syn030}):}
It is shown that when memory size is up to 4MB, \aname{} achieves a comparable ARE with Elastic in two datasets, and is $1.63 \sim 2.75$ times lower than CU, $2.78 \sim 5.23$ times lower than CM, $2.76 \sim 5.05$ times lower than Coco, and $18.07 \sim 36.67$ times lower than Space-Saving.

\presub
\subsection{Speed Comparison.} \postsub
\label{subsec:exp:Speedcompare}

\begin{figure}[htbp]
    \centering
	\includegraphics[width=0.85\linewidth, ]{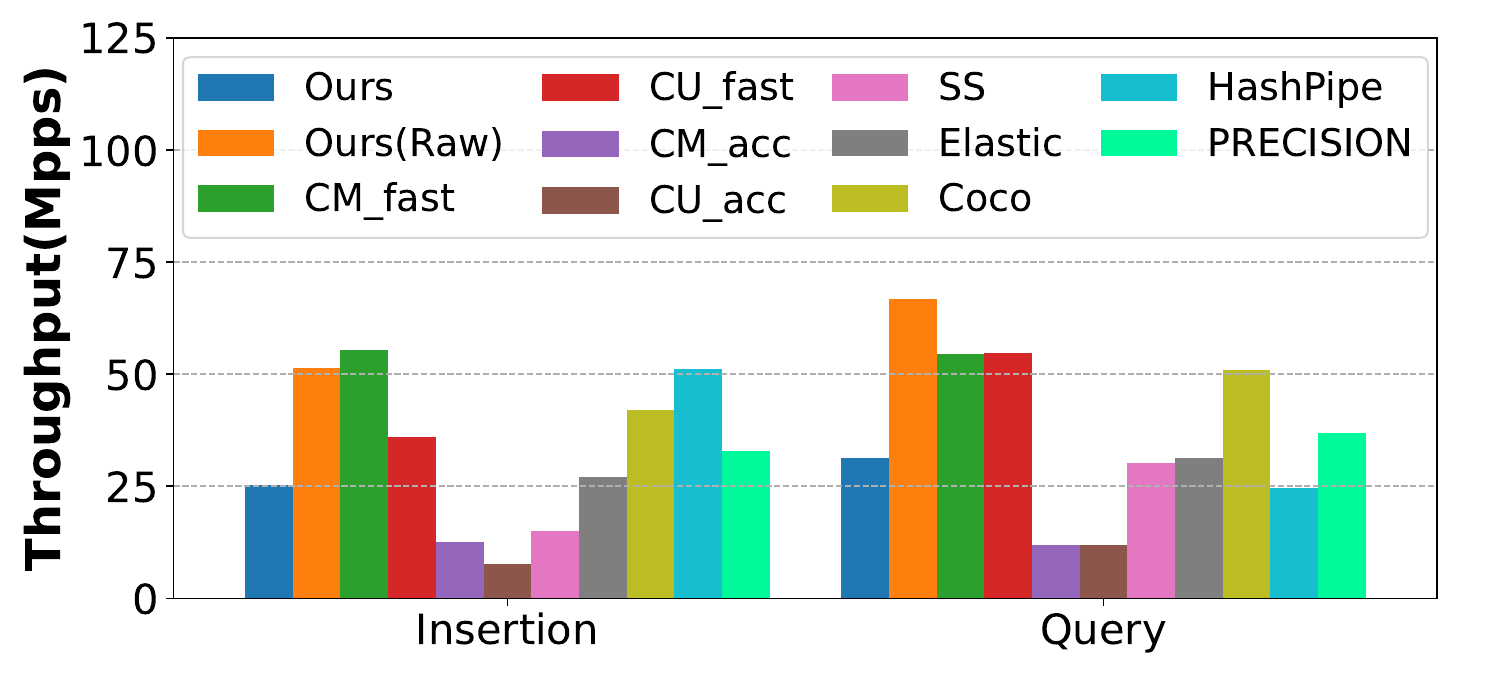}
    \prefigcaption
    \caption{Throughput Evaluation.}
	\label{fig:app:tp}
\postfig
\end{figure}

We find that \anamelong{} is not only highly accurate but also fast. In our tests involving 10 million insertions and queries, we compare its throughput with that of other algorithms. An alternative version of \anamelong{}, without the mice filter (``Raw'' in the figure), is also presented, sacrificing a tolerable level of accuracy for a significant increase in speed.

Figure \ref{fig:app:tp} shows that the insertion throughputs for \anamelong{} and the Raw version are 25.40 Mpps and 51.29 Mpps, respectively. The Raw version's throughput is comparable to CM (fast), Coco, and HashPipe, and it is about 1.42 times faster than CU (fast), 1.43 times faster than Elastic, and 1.56 times faster than PRECISION, surpassing other algorithms by 3.40 to 6.65 times. For query throughput, \anamelong{} and its Raw variant achieve 31.29 Mpps and 66.89 Mpps, respectively. The Raw version is around 1.22 times quicker than CM (fast), CU (fast), and Coco, 1.81 times faster than PRECISION, 2.14 times faster than Elastic, and 2.22 to 5.62 times faster than others.




\presub
\subsection{Impact of Parameters}
\postsub
\label{subsec:exp:Parameters}

We explore the impact of various parameters on the accuracy of \anamelong{}, including \(R_w\), \(R_\lambda\), and the error threshold \(\Lambda\), and analyze the trends in \anamelong{}'s speed changes. At the same time, we provide recommended parameter settings.


\subsubsection{Impact of Parameter $R_w$. }
\label{subsubsec:exp:r_w}

When adjusting $R_w$ (\ie, the parameter of the decreasing speed of array size), we find \aname{} performs best when $R_w=2$.

\begin{figure}[htbp]
\prefig
    \centering
 	    \begin{subfigure}{0.47\linewidth}
        \centering
        \includegraphics[width=\textwidth]{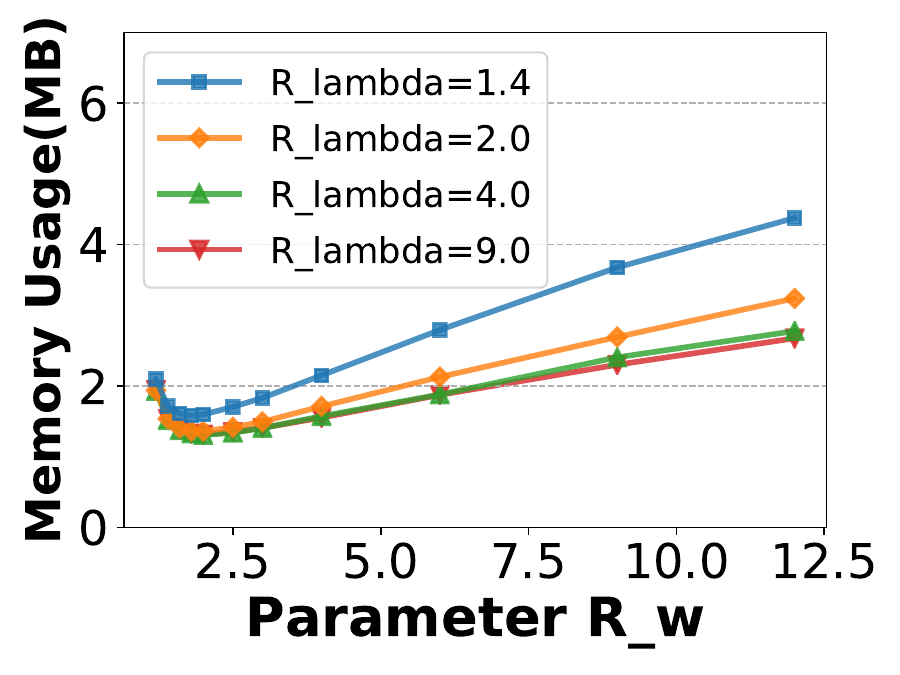}
        \presubfigcaption
        \caption{IP Trace}
        \label{subfig:nature:rw_range_caida}
        \end{subfigure}
    \hfill
 	    \begin{subfigure}{0.47\linewidth}
        \centering
        \includegraphics[width=\textwidth]{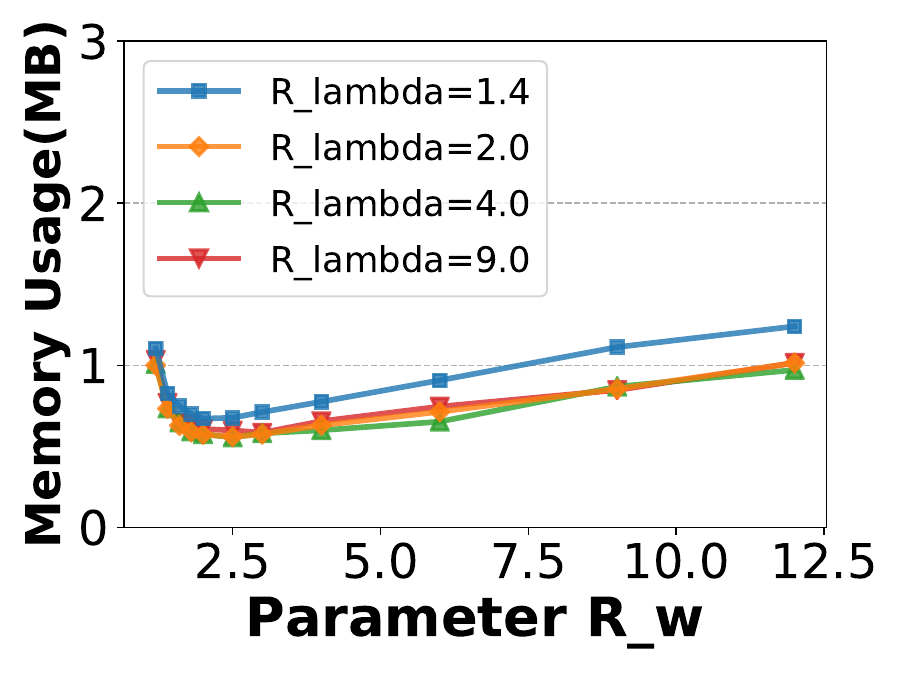}
        \presubfigcaption
        \caption{Web Stream}
        \label{subfig:nature:rw_range_webpage}
        \end{subfigure}
        
    \prefigcaption
    \caption{{Impact of $R_w$ under Zero Outlier.}}
    \label{fig:nature:rw_range}
\postfig
\end{figure}

\ppp{Memory Usage under Zero Outlier (Figure  \ref{subfig:nature:rw_range_caida}, \ref{subfig:nature:rw_range_webpage}):} 
We conduct experiments on IP Trace and Web Stream datasets. We set the user-defined threshold $\Lambda$ to 25, and compare the minimum memory consumption achieving zero outlier under different $R_w$.
It is shown that \aname{} with $R_w=2 \sim 2.5$ requires less memory than the ones with other $R_w$. When $R_w$ is lower than 1.6 or higher than 3, the memory consumption increases rapidly.

\begin{figure}[htbp]
\prefig
    \centering
 	    \begin{subfigure}{0.47\linewidth}
        \centering
        \includegraphics[width=\textwidth]{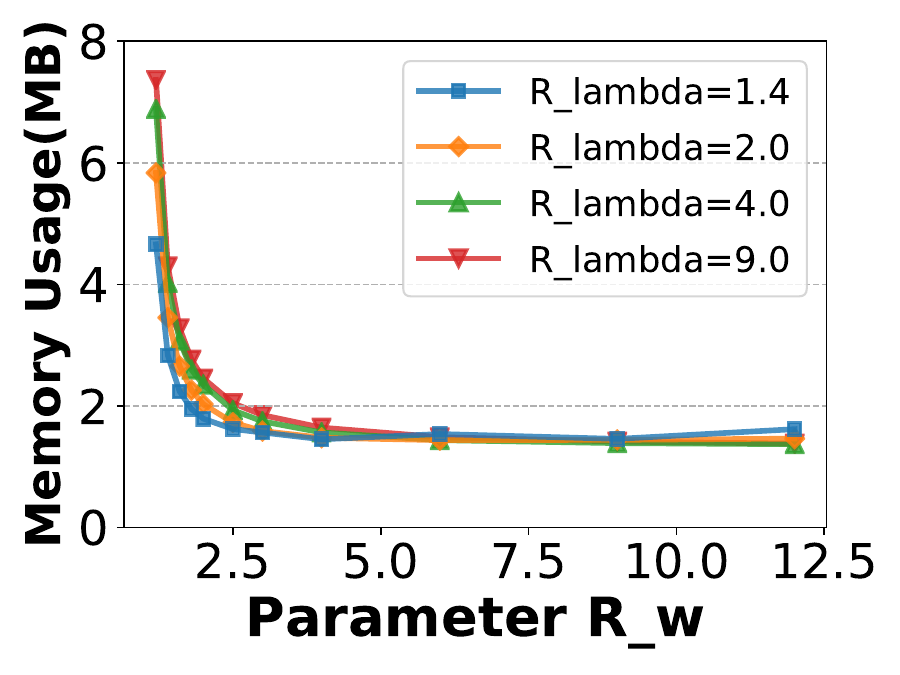}
        \presubfigcaption
        \caption{IP Trace}
        \label{subfig:nature:rw_freq_caida}
        \end{subfigure}
    \hfill
 	    \begin{subfigure}{0.47\linewidth}
        \centering
        \includegraphics[width=\textwidth]{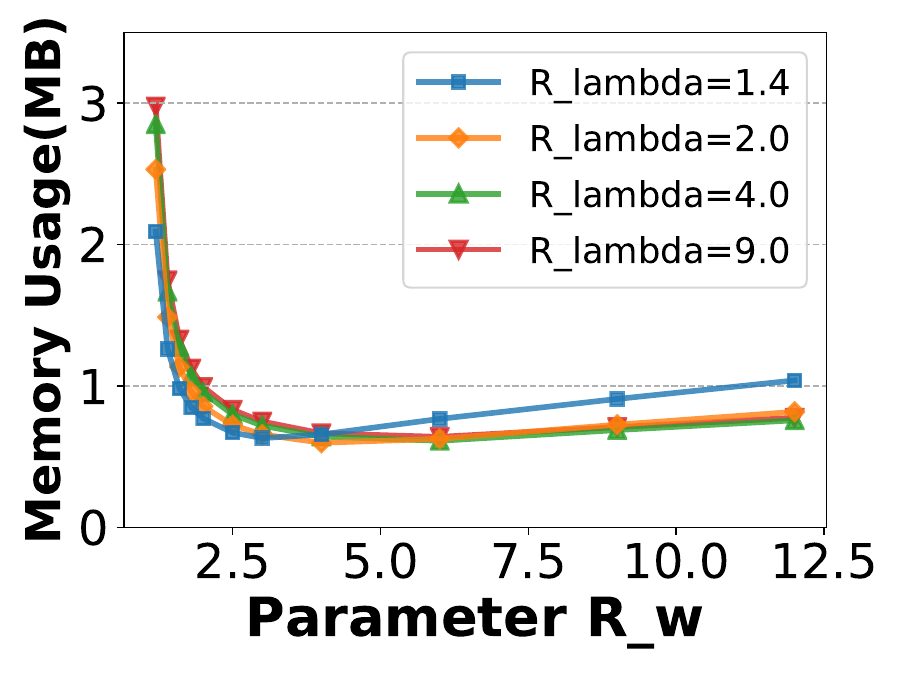}
        \presubfigcaption
        \caption{Web Stream}
        \label{subfig:nature:rw_freq_webpage}
        \end{subfigure}
        
    \prefigcaption
    \caption{{Impact of $R_w$ under the Same Average Error.}}
    \label{fig:nature:rw_freq}
\postfig
\end{figure}

\ppp{Memory Usage under the Same Average Error (Figure  \ref{subfig:nature:rw_freq_caida}, \ref{subfig:nature:rw_freq_webpage}):} 
We conduct experiments on IP Trace and Web Stream datasets, set the target estimation AAE to 5, and compare the memory consumption when $R_w$ varies. The figures show that the higher $R_w$ goes with less memory usage. However, the memory consumption is quite close to the minimum value when $R_w= 2 \sim 6$. 

\subsubsection{Impact of Parameter $R_\lambda$. }
\label{subsubsec:exp:r_l}
When adjusting $R_\lambda$ (\ie, the parameter of the decreasing speed of error threshold), we find \aname{} performs best when we set $R_\lambda=2.5$.

\begin{figure}[htbp]
\prefig
    \centering
 	    \begin{subfigure}{0.47\linewidth}
        \centering
        \includegraphics[width=\textwidth]{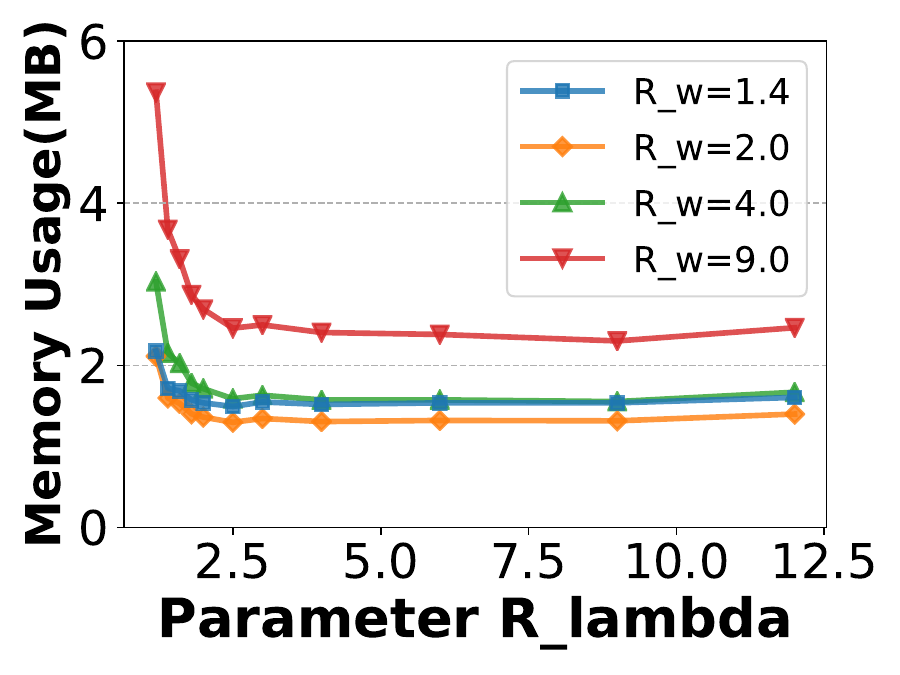}
        \presubfigcaption
        \caption{IP Trace}
        \label{subfig:nature:rl_range_caida}
        \end{subfigure}
    \hfill
 	    \begin{subfigure}{0.47\linewidth}
        \centering
        \includegraphics[width=\textwidth]{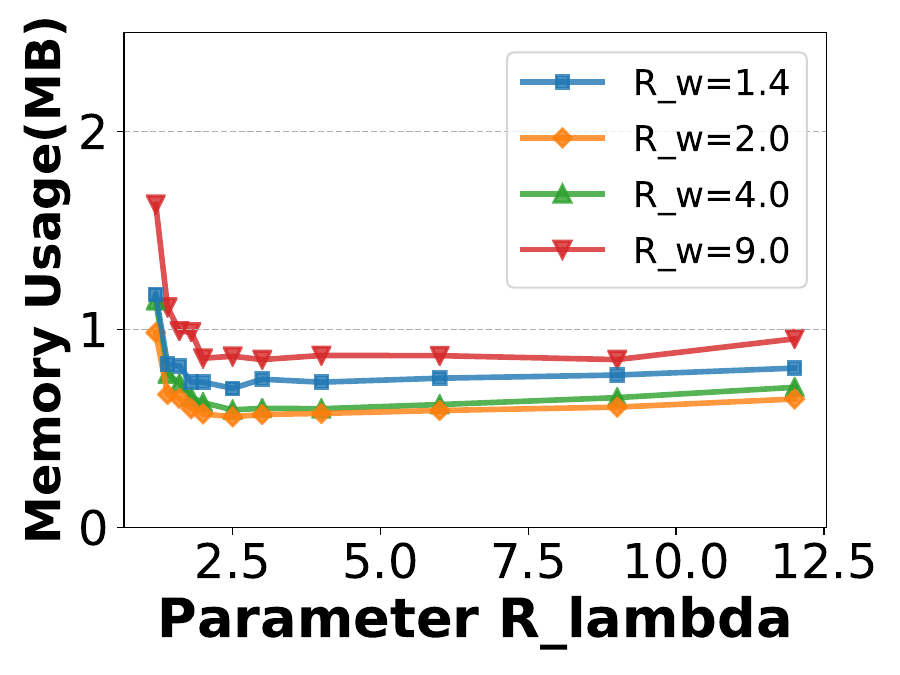}
        \presubfigcaption
        \caption{Web Stream}
        \label{subfig:nature:rl_range_webpage}
        \end{subfigure}
        
    \prefigcaption
    \caption{{Impact of $R_\lambda$ under Zero Outlier.}}
    \label{fig:nature:rl_range}
\postfig
\end{figure}

\ppp{Memory Usage under Zero Outlier (Figure  \ref{subfig:nature:rl_range_caida}, \ref{subfig:nature:rl_range_webpage}):} 
We conduct experiments and set the target $\Lambda$ to 25. As the figures show, memory consumption drops down rapidly when $R_\lambda$ grows from 1.2 to 2 and finally achieve the minimum when $R_\lambda = 2$. There is no significant change when $R_\lambda$ is higher than 2.5, only some jitters due to the randomness of \aname{}. 

\begin{figure}[htbp]
\prefig
    \centering
 	    \begin{subfigure}{0.47\linewidth}
        \centering
        \includegraphics[width=\textwidth]{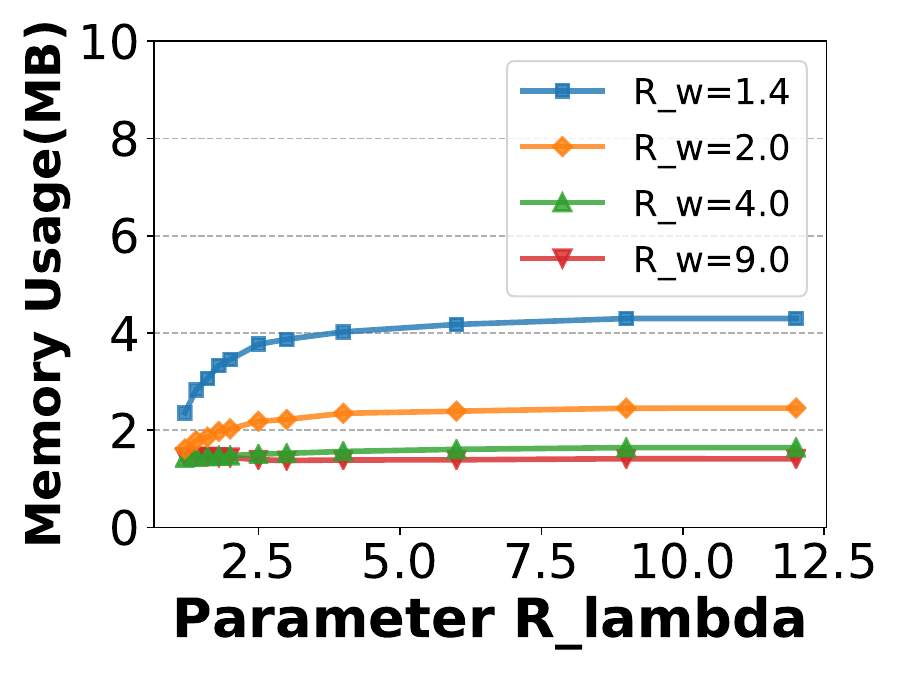}
        \presubfigcaption
        \caption{IP Trace}
        \label{subfig:nature:rl_freq_caida}
        \end{subfigure}
    \hfill
 	    \begin{subfigure}{0.47\linewidth}
        \centering
        \includegraphics[width=\textwidth]{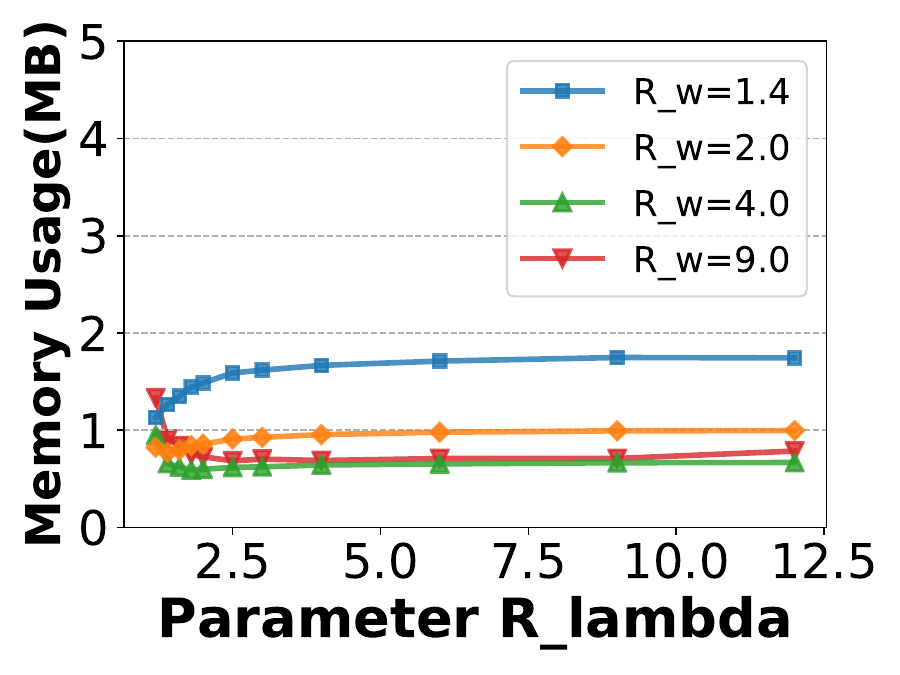}
        \presubfigcaption
        \caption{Web Stream}
        \label{subfig:nature:rl_freq_webpage}
        \end{subfigure}
        
    \prefigcaption
    \caption{{Impact of $R_\lambda$ under the Same Average Error.}}
    \label{fig:nature:rl_freq}
\postfig
\end{figure}

\ppp{Memory Usage under the Same Average Error (Figure  \ref{subfig:nature:rl_freq_caida}, \ref{subfig:nature:rl_freq_webpage}):} 
To explore the impact of parameter $R_\lambda$ on average error, we set the target estimation AAE to 5 and compare the memory consumption when $R_\lambda$ varies. It is shown that when $R_w$ is low, the higher $R_\lambda$ is, the less memory \aname{} uses. When $R_w$ is greater than 4, $R_\lambda$ affects little.

\subsubsection{Error Threshold $\Lambda$. }
\label{subsubsec:exp:eps}

We find that the user-defined error threshold $\Lambda$, which denotes the maximum estimated error \aname{} guaranteed, is almost inversely proportional to the memory consumption. 

\begin{figure}[htbp]
\prefig
    \centering
 	    \begin{subfigure}{0.47\linewidth}
        \centering
        \includegraphics[width=\textwidth]{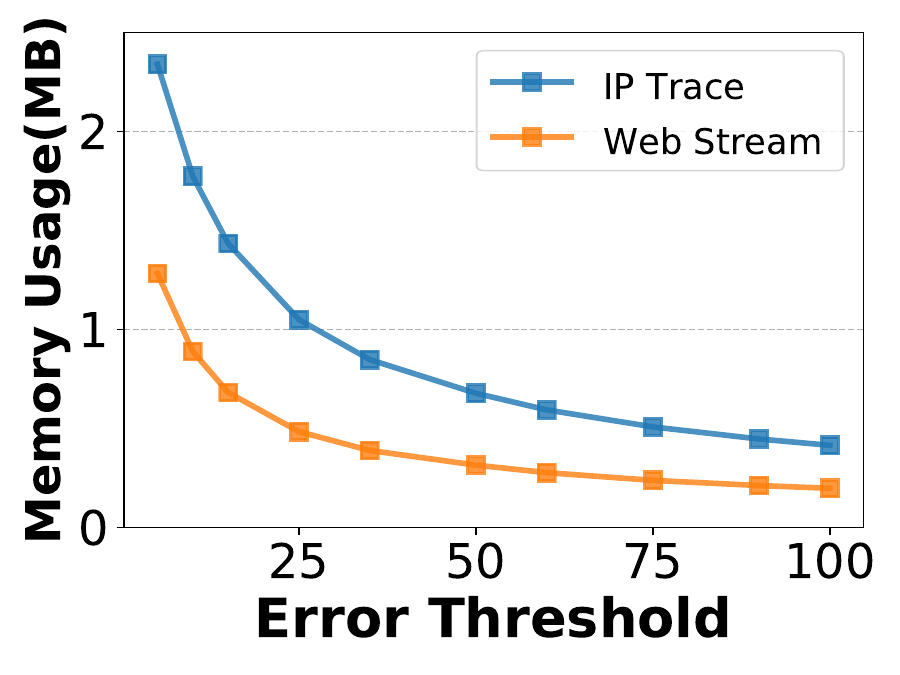}
        \presubfigcaption
        \caption{Zero Outlier}
        \label{subfig:nature:eps_range_caida}
        \end{subfigure}
    \hfill
 	    \begin{subfigure}{0.47\linewidth}
        \centering
        \includegraphics[width=\textwidth]{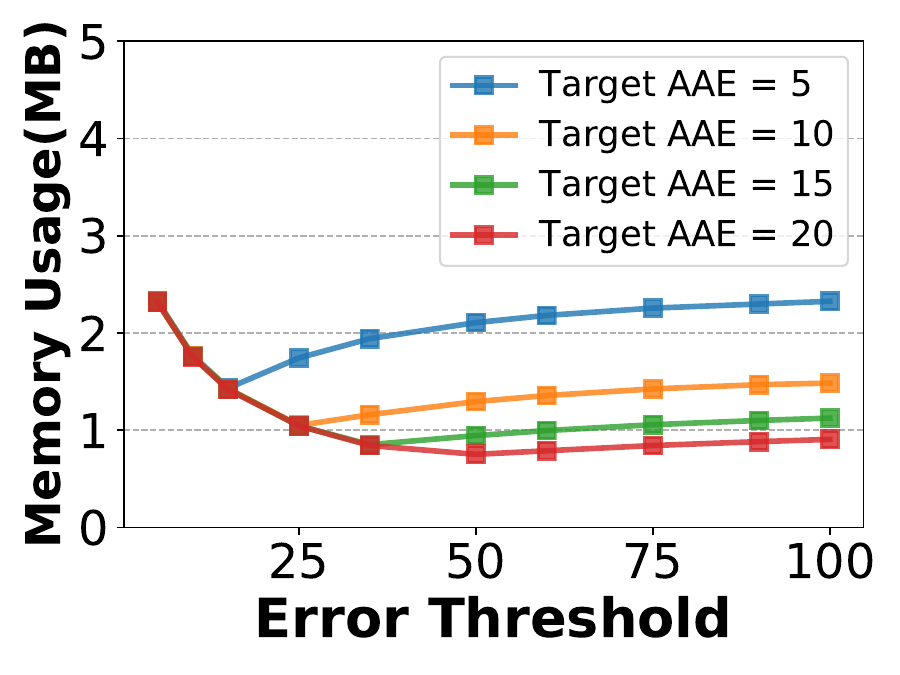}
        \presubfigcaption
        \caption{Overall AAE=5}
        \label{subfig:nature:eps_freq_caida}
        \end{subfigure}
        
    \prefigcaption
    \caption{{Memory Usage for Different $\Lambda$.}}
    \label{fig:nature:eps_range}
\postfig
\end{figure}

\ppp{Memory Usage under Zero Outlier (Figure \ref{subfig:nature:eps_range_caida}):} In this experiment, we fix the parameter $R_w$ to 2,  $R_\lambda$ to 2.5, and conduct it on three different datasets. 
It is shown that memory usage monotonically decreases, which means the optimal $\Lambda$ is exactly the maximum tolerable error.
On the other hand, reducing $\Lambda$ blindly will lead to an extremely high memory cost.

\ppp{Memory Usage under the Same Average Error (Figure \ref{subfig:nature:eps_freq_caida}):} In this experiment, we fix the parameter $R_w$ to 2,  $R_\lambda$ to 2.5, and conduct it on IP Trace dataset. The figure shows that optimal $\Lambda$ increases as target AAE increases, and optimal $\Lambda$ is about 2 $\sim$ 3 times greater than target AAE. For target AAE=5/10/15/20, the optimums are 15/25/35/50, requiring 1.43/1.05/0.85/0.75MB memory respectively.

\subsubsection{Trend of speed changes. }
\label{subsubsec:exp:speed}

\begin{figure}[htbp]
\prefig
    \centering
 	    \begin{subfigure}{0.47\linewidth}
        \centering
		\includegraphics[width=\textwidth, ]{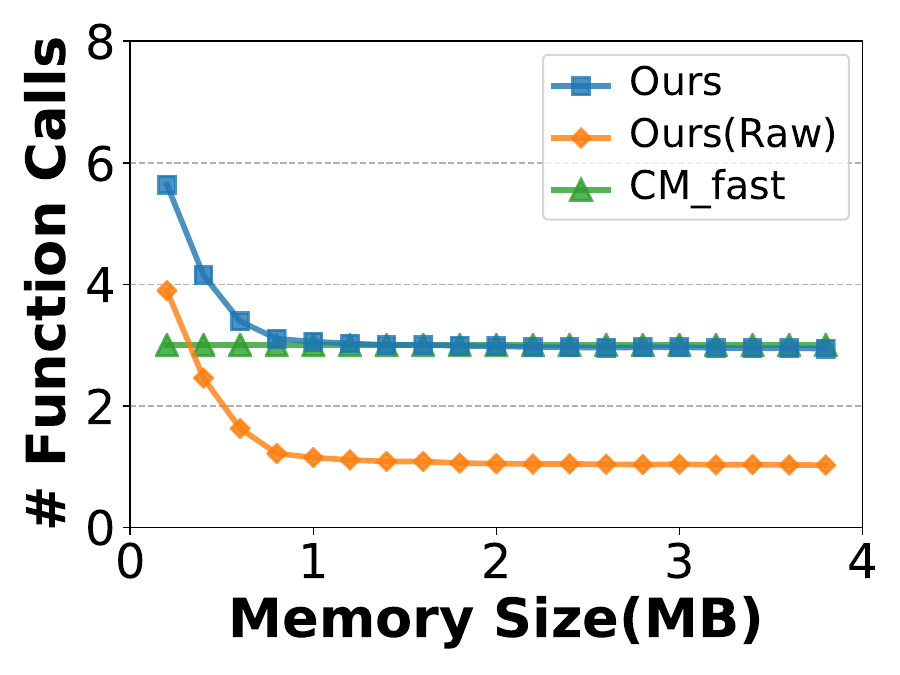}
        \presubfigcaption
        \caption{Item Insertion}
		\label{subfig:tp:hash_insert}
        \end{subfigure}
    \hfill
 	    \begin{subfigure}{0.47\linewidth}
        \centering
		\includegraphics[width=\textwidth, ]{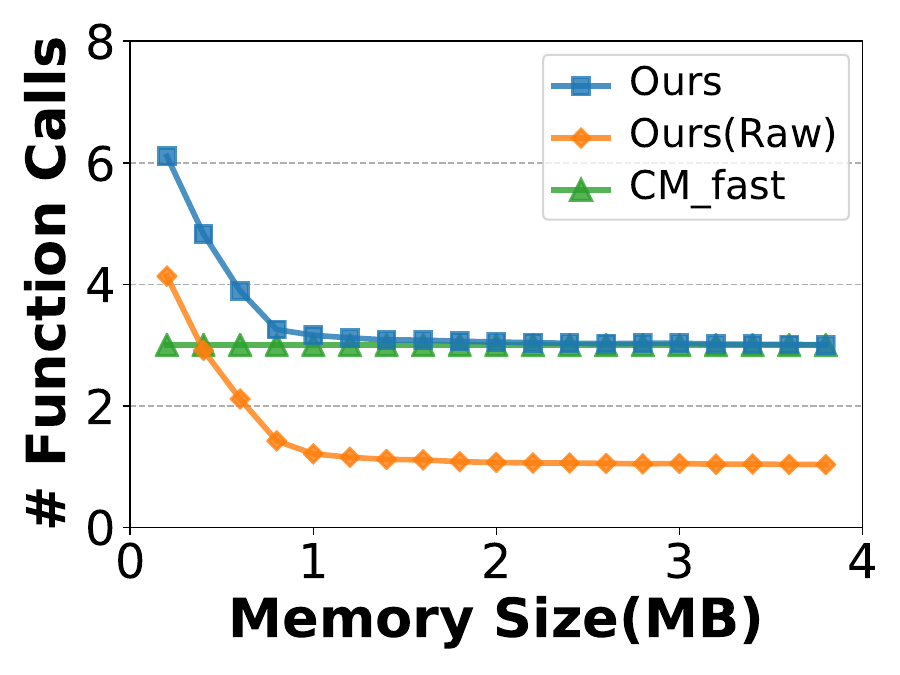}
        \presubfigcaption
        \caption{Frequency Query}
		\label{subfig:tp:hash_query}
        \end{subfigure}
    \prefigcaption
    \caption{Average Number of Hash Function Calls.}
	\label{fig:tp:hash}
\postfig
\end{figure}
The number of hash function calls, directly proportional to consumed time, fundamentally indicates the trend of speed changes. 
In \anamelong{}, this number dynamically varies during insertions and queries due to its multi-layer structure. To explore the relationship between memory size and the average number of hash function calls, we conducted experiments using the IP Trace dataset.

Figures \ref{subfig:tp:hash_insert} and \ref{subfig:tp:hash_query} reveal that the average hash function calls for the raw version of \anamelong{} decrease rapidly with increasing memory, eventually stabilizing at 1. \anamelong{} with a 2-array mice filter eventually stabilizes at 3 due to 2 additional calls in the filter. Smaller \anamelong{} instances record fewer keys in the earlier layers, leading to more hash function calls and reduced throughput. For this reason, unless memory is exceptionally scarce, we recommend allocating more space to gain faster processing speeds.

\presub
\subsection{In-depth Observations of \aname{}} \postsub
\label{subsec:exp:Observations}

We show how \aname{} performs in \goal{} comprehensively, and compare it with prior algorithms. 

\presubsub
\subsubsection{Error-Sensing Ability. }
\label{subsubsec:exp:sense}
\anamelong{} can confidently and accurately sense the error, the MPE it reports, using the default parameters.

\begin{figure}[htbp]
\prefig
    \centering
 	    \begin{subfigure}{0.47\linewidth}
        \centering
		\includegraphics[width=\textwidth, ]{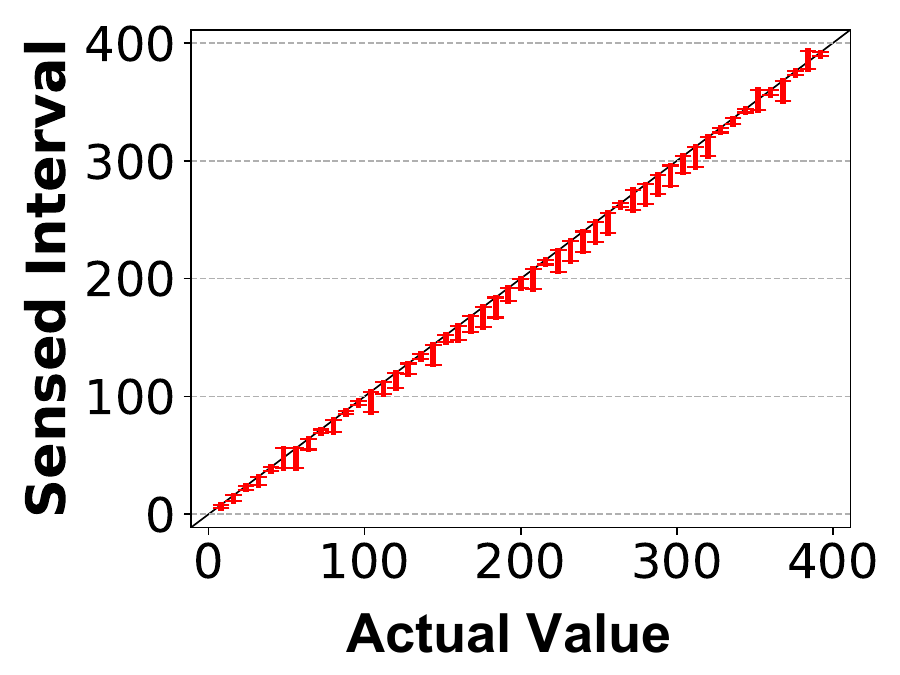}
        \presubfigcaption
        \caption{Mice Keys}
		\label{subfig:nature:7.2.1.1}
        \end{subfigure}
    \centering
 	    \begin{subfigure}{0.47\linewidth}
        \centering
		\includegraphics[width=\textwidth, ]{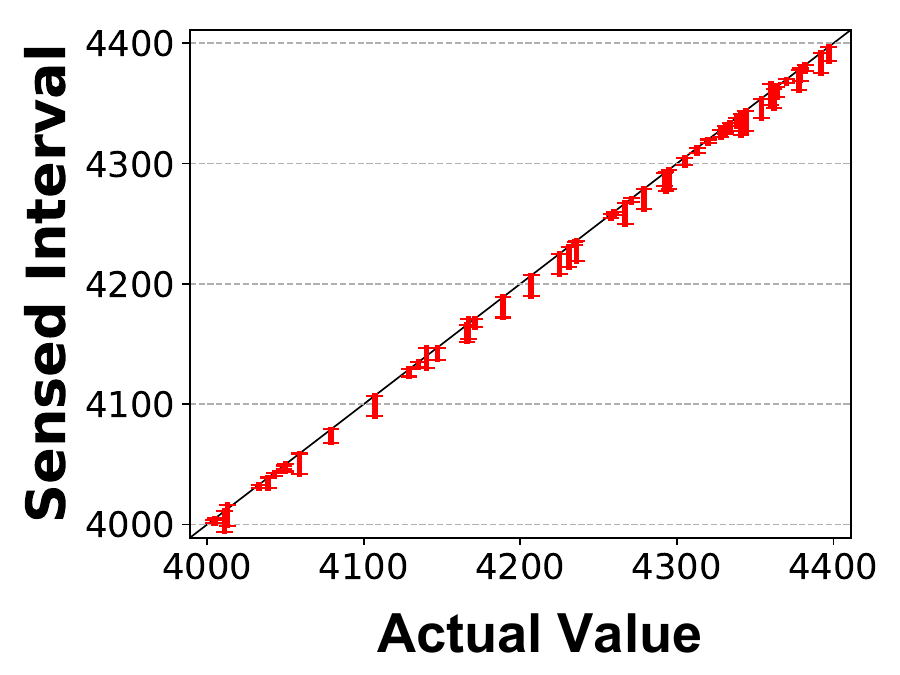}
        \presubfigcaption
        \caption{Elephant Keys}
		\label{subfig:nature:7.2.1.1_2}
        \end{subfigure}
    \prefigcaption
    \caption{Illustration of Sensed Error and Intervals.}
	\label{fig:nature:overview}
\postfig
\end{figure}

\ppp{Sensed Interval (Figure  \ref{subfig:nature:7.2.1.1}, \ref{subfig:nature:7.2.1.1_2}):}
We examine keys with both large and small values to ensure their true values fall within the range [estimated value - MPE, estimated value], thereby confirming the correctness.


\begin{figure}[htbp]
\prefig
    \centering
 	    \begin{subfigure}{0.47\linewidth}
        \centering
		\includegraphics[width=\textwidth, ]{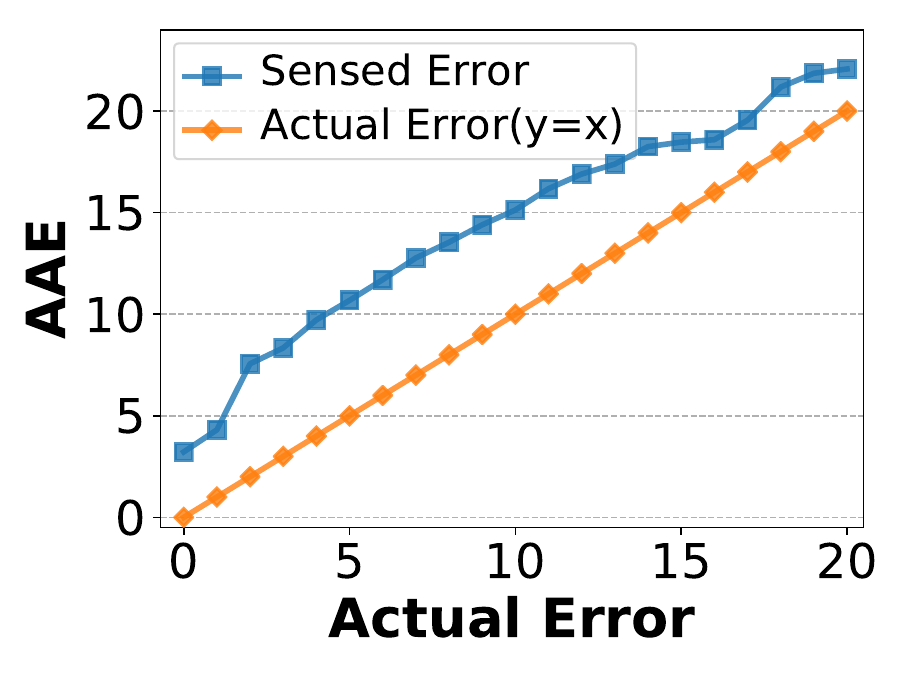}
        \presubfigcaption
        \caption{vs. Actual Error}
		\label{subfig:nature:7.2.1.2}
        \end{subfigure}
    \centering
 	    \begin{subfigure}{0.47\linewidth}
        \centering
		\includegraphics[width=\textwidth, ]{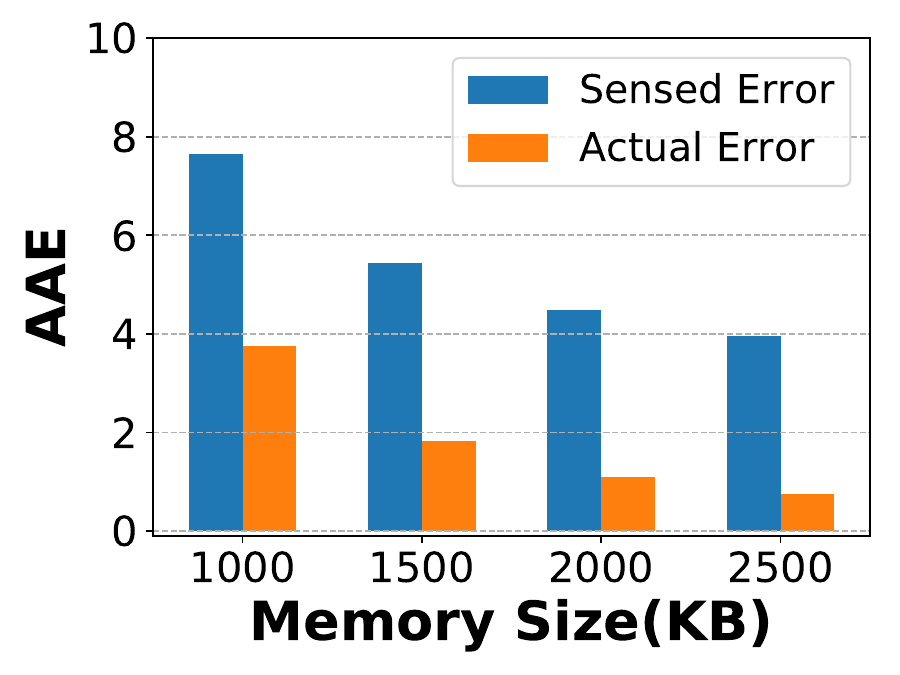}
        \presubfigcaption
        \caption{vs. Memory Size}
		\label{subfig:nature:7.2.1.3}
        \end{subfigure}
    \prefigcaption
    \caption{Experiments on Sensed Error.}
	\label{fig:nature:sensed_error}
\postfig
\end{figure}

\ppp{Actual Error \emph{vs.} Sensed Error (Figure  \ref{subfig:nature:7.2.1.2}):}
As we query the values of all keys in \aname{}, we classify these keys by their actual absolute error, and calculate the average sensed error respectively.
The result shows that the average sensed error keeps close to the actual error no matter how it changes, which means \aname{} can sense error accurately and stably.

\ppp{Sensed Error \emph{vs.} Memory Size  (Figure  \ref{subfig:nature:7.2.1.3}):}
We further vary the memory size from 1000KB to 2500KB, and study how errors change.
The figure shows that sensed error decreased when memory grows.


\presubsub
\subsubsection{Error-Controlling Ability. }
\label{subsubsec:exp:control}
\aname{} controls error efficiently as our expectation.
 

\begin{figure}[htbp]
    \centering
 	    \begin{subfigure}{0.47\linewidth}
        \centering
		\includegraphics[width=\textwidth, ]{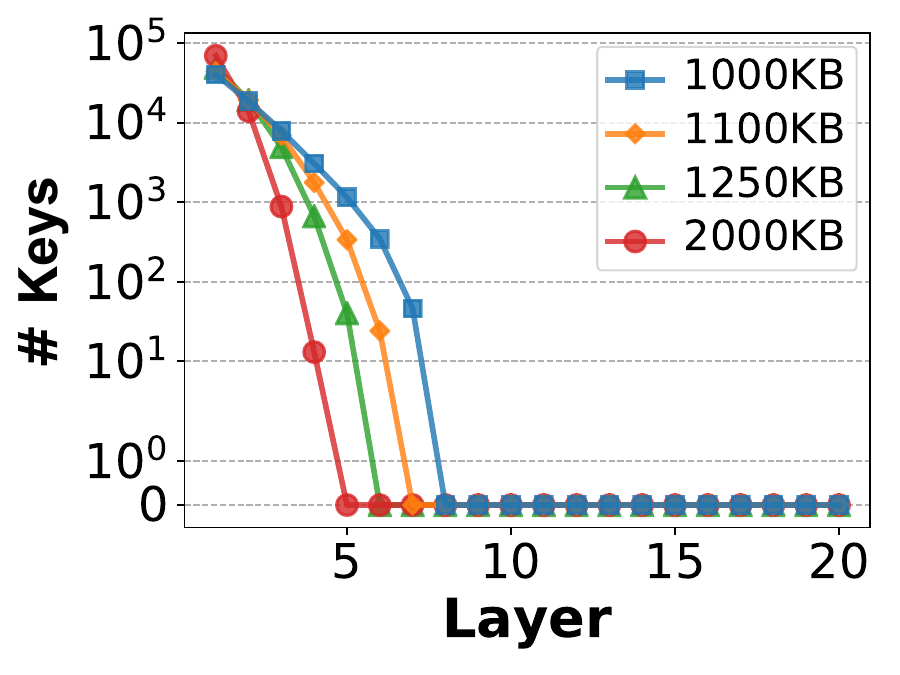}
        \presubfigcaption
        \caption{Layer Distribution }
		\label{subfig:nature:7.2.2.1}
        \end{subfigure}
    \hfill
 	    \begin{subfigure}{0.47\linewidth}
        \centering
		\includegraphics[width=\textwidth, ]{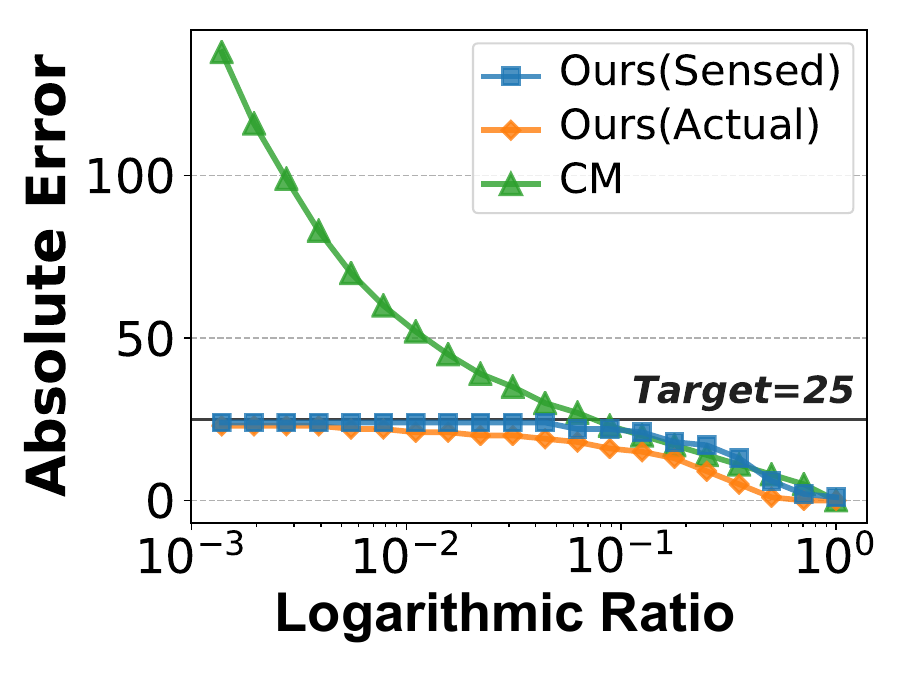}
        \presubfigcaption
        \caption{Error Distribution}
		\label{subfig:nature:7.2.2.2}
        \end{subfigure}
        
    \prefigcaption
    \caption{ Illustration of Error-Controlling.}
    \vspace{-0.1in}
	\label{fig:nature:error_control}
\postfig
\end{figure}

\ppp{Layer Distribution (Figure  \ref{subfig:nature:7.2.2.1}):}
When the latest-arriving item of a key concludes its insertion in a particular layer, we categorize the key as belonging to that layer. Through repeated experiments, we calculate the distribution of keys across layers. The results, as depicted in the figure, indicate that the number of keys associated with each layer diminishes at a rate faster than exponential. This suggests that \anamelong{} is capable of effectively controlling errors with only a few layers, and the remaining layers contribute to eliminating potential outliers.

\ppp{Error Distribution (Figure  \ref{subfig:nature:7.2.2.2}):}
We count absolute errors of all keys, and sort them in descending order. 
The figure shows that errors of \aname{} are controlled within $\Lambda$ completely, while most traditional sketch algorithms cannot control the error of all keys, such as CM.



\presubsub
\subsubsection{Deployment}
\label{subsubsec:exp:control}
\begin{figure}[htbp]
\prefig
    \centering
 	    \begin{subfigure}{0.47\linewidth}
        \centering
		\includegraphics[width=\textwidth, ]{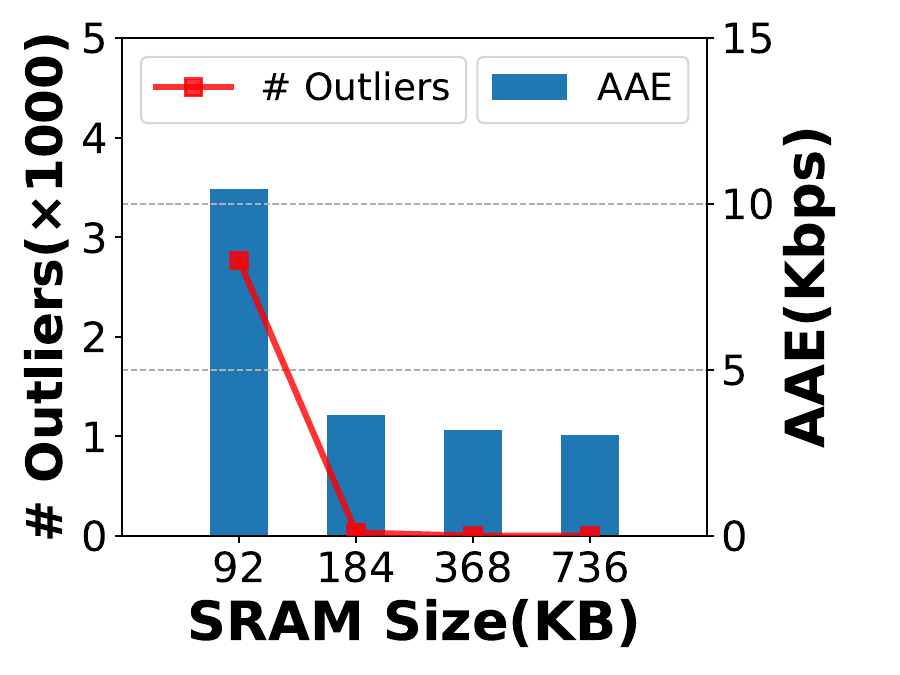}
        \presubfigcaption
        \caption{IP Trace}
		\label{subfig:testbed:acc:caida}
        \end{subfigure}
 	    \begin{subfigure}{0.47\linewidth}
        \centering
		\includegraphics[width=\textwidth, ]{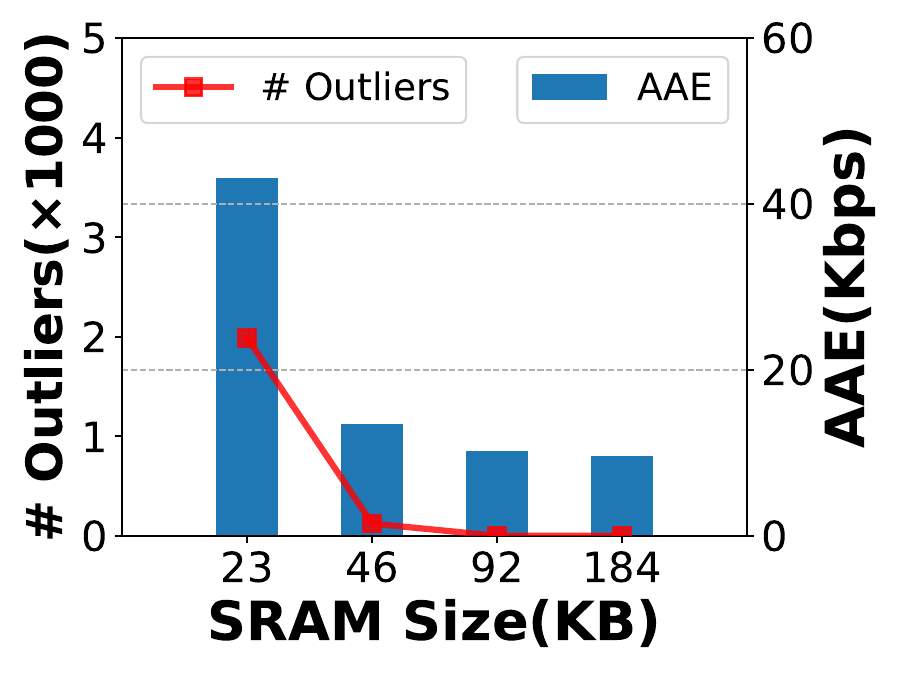}
        \presubfigcaption
        \caption{Hadoop}
		\label{subfig:testbed:hadoop}
        \end{subfigure}
        
    \prefigcaption
    \caption{Accuracy on TestBed Deployment.}
	\label{fig:testbed:acc}
\postfig
\end{figure}

Compared to the flexible CPU platforms, high-performance devices impose more restrictions on algorithm implementation.
The more complex the operations used by an algorithm, the less likely it is to be feasibly deployed in reality. 
We evaluate the accuracy of \anamelong{} implemented on a programmable switch known as Tofino. We send 40 million packets selected from the IP Trace and Hadoop datasets at a link speed of 40Gbps from an end-host connected to the Tofino switch. The evaluation focuses on AAE and the number of outliers for \anamelong{} using SRAM of different sizes.

As depicted in Figure \ref{fig:testbed:acc}, for the IP Trace dataset, \anamelong{} requires more than 368KB of SRAM to ensure zero outliers, maintaining an AAE within 4Kbps. For the Hadoop dataset, more than 92KB of SRAM is necessary for \anamelong{} to guarantee no outliers, with an AAE within 10Kbps.

\presec
\section{Related Work} \postsec
\label{sec:relatedend}

Sketches, as a type of probabilistic data structure, have garnered significant interest due to their diverse capabilities. They are most often used for addressing stream summary problem. 
When all values in the stream equal one, the problem is also known as the frequency estimation. 
We categorize existing sketches into Counter-based and Heap-based, and supplement the related work not detailed previously.
\ppp{Counter-based Sketches} are composed of counters, including CM \cite{cmsketch}, CU \cite{estan2002cusketch}, Count \cite{countsketch}, Elastic \cite{yang2018elastic}, UnivMon \cite{liu2016UnivMon}, Coco \cite{zhang2021cocosketch}, SALSA \cite{basat2021salsa}, DHS \cite{zhao2021dhs}, SSVS \cite{melissourgossingle} and more \cite{namkung2022sketchlib, nsdi2022HeteroSketch, huang2017sketchvisor, huang2021toward}.
Among them, the work most relevant to \aname{} is the Elastic \cite{yang2018elastic}, which likewise employs election, with two counters resembling YES and NO. But because Elastic's purpose is to find frequent keys, it resets the NO-like counter to 1 when a replacement occurs, making it incapable of sensing errors. \aname{} and Elastic are similar only in appearance, as they differ greatly in their target problems and underlying ideas.

\ppp{Heap-based Sketches} include Frequent \cite{frequent1}, Space Saving \cite{spacesaving}, Unbiased Space Saving \cite{unbiasedsketch}, SpaceSaving$^{\pm}$ \cite{zhao2021spacesaving} and more \cite{boyer1991majority}. 
The insertions of these solutions rely on a heap structure, resulting in slower speeds. Compared to their logarithmic time complexity, \aname{} achieves an amortized complexity of \(O(1+\Delta\ln\ln(\frac{N}{\Lambda}))\). Heap-based Sketches can only be optimized using a linked list in the special case where the value equals 1, achieving an insertion efficiency of O(1). However, even in the scenario where the value is 1, \aname{} still retains its unique advantages, being more suitable for high-performance hardware implementation, while heaps or linked lists are too hard to be implemented \cite{basat2020Precision}.
 

\ppp{Other Sketches.}
Besides stream summary, sketches can address various tasks, including estimating cardinality \cite{hyperloglog, linear-counting}, quantiles \cite{karnin2016optimalKLL,zhao2021kll,masson12ddsketch,shahout2023togetherSQUAD}, and join sizes \cite{wang2023joinsketch,cormode2005sketching,alon1999tracking,alon1996space}, among other tasks \cite{sketchsurvey}. 

	
\presec
\section{Conclusion} \postsec
\label{sec:conclusion}

We focus on approximating sums of values in data streams for all keys with a high degree of confidence, aiming to prevent the occurrence of outliers with excessive errors. To this end, we have developed ReliableSketch, which provides high-confidence guarantees for all keys. It features near-optimal amortized insertion time of \(O(1 + \Delta\ln\ln(\frac{N}{\Lambda}))\), near-optimal space complexity \(O(\frac{N}{\Lambda} + \ln(\frac{1}{\Delta}))\), and excellent hardware compatibility. Compared to counter-based sketches, ReliableSketch optimizes confidence, speed, and space usage; and against heap-based sketches, it offers superior speed and, despite theoretically larger space requirements, demonstrates more efficient space utilization in practice.

The key idea of ReliableSketch is to identify keys with significant errors and effectively control these errors to completely eliminate outliers. These two steps are facilitated by our key techniques, ``Error-Sensible Bucket'' and ``Double Exponential Control''. We have implemented ReliableSketch on CPU servers, FPGAs, and programmable switches. Our experiments indicate that under the same limited space, ReliableSketch not only maintains errors for all keys below \(\Lambda\) but also achieves near-optimal throughput, surpassing competitors that struggle with thousands of outliers.

We have made our source code publicly available.

	\vfill\eject

    \clearpage	
 	\bibliographystyle{unsrt}
	\bibliography{main}	


	\vfill\eject
\appendix

{\centering \section*{APPENDIX}}

\presec
\section{Mathematical Proofs} \postsec
\label{sec:appendix:math}

\renewcommand{\thetheorem}{\Alph{section}.\arabic{theorem}}
\setcounter{theorem}{0}

\subsection{Preliminaries}
\postsub
\label{app:math:pre}

We derive a lemma to bound the sum of $n$ random variables. This lemma is similar to the Hoeffding bound but cannot be replaced by Hoeffding.

\begin{lemma}
\label{app:basic}
Let $X_1,\dots, X_n$ be $n$ random variables such that
\begin{align*}
    X_i\in\{0,s_i\}, &&
    \Pr(X_i=s_i~|~X_1,\cdots,X_{i-1})\leqslant p,
\end{align*}
where $0\leqslant s_i \leqslant 1$.
Let $X=\sum_{i=1}^n X_i$, and $\mu=\sum_{i=1}^n ps_i=nmp$.
\begin{align*}
    Pr\left(X>(1+\Delta)\mu\right) \leqslant e^{-(\Delta-(e-2))nmp}
\end{align*}
\end{lemma}

\begin{proof}
For any $t > 0$, by using the Markov inequality we have
\begin{align*}
    \Pr(X>(1+\Delta)\mu) 
    =
    \Pr(e^{X}>e^{(1+\Delta)\mu}) 
    \leqslant
    \frac{E(e^{X})}{e^{(1+\Delta)\mu}}.
\end{align*}
According to the conditions, we have
\begin{align*}
    &
    E(e^{X}) 
    = 
    E\left(E(e^{\sum_{i=1}^n{X_i}}~|~X_1,\cdots,X_{n-1})\right)
    \\
    =&
    E\left(
    \begin{aligned}
    e^{\sum_{i=1}^{n-1}{X_i}}
    \cdot \Pr(X_n=0~|~X_1,\cdots,X_{n-1})
    \\+
    e^{s_n+\sum_{i=1}^{n-1}{X_i}}
    \cdot \Pr(X_n=s_n~|~X_1,\cdots,X_{n-1})
    \end{aligned}
    \right)
    \\
    \leqslant&
    E\left(
    e^{\sum_{i=1}^{n-1}{X_i}}
    \right)
    \cdot
    \left(1+p\left(e^{s_n}-1\right)\right)
    \leqslant
    \cdots
    \leqslant 
    \prod_{i=1}^n  \left(1+p\left(e^{s_i}-1\right)\right)
\end{align*}
Because of $1+x < e ^x$, we have
\begin{align*}
    E(e^{X})
    \leqslant
    \prod_{i=1}^n e^{p\left(e^{s_i}-1\right)}.
\end{align*}
Since for $s_i\leqslant 1$, there is $e^{s_i}-1 \leqslant (e-1)s_i$, so there is
\begin{align*}
    E(e^{X})
    \leqslant
    e^{\sum_{i=1}^n p(e-1) s_i} = e^{(e-1)mnp}.
\end{align*}
That is
\begin{align*}
    Pr(X>(1+\Delta)\mu)
    \leqslant
    \frac{e^{(e-1)nmp}}{e^{(1+\Delta)nmp}}
    =
    e^{-(\Delta-(e-2))nmp}
\end{align*}

\end{proof}

\subsection{Definition of Symbols}
\label{app:math:def}

\begin{enumerate}[leftmargin=*]
    \setlength{\itemsep}{0.2cm}
    
    \item $\mathcal{S}_{i}$: $\{e_1,\cdots,e_{N_i}\}$, the set of keys entering the $i$-th layer, where $N_i=|\mathcal{S}_i|$.

    \item $f_i(e)$: the number of times that key $e$ enters the $i$-th layer.
    
    \item $\mathcal{S}^{0}_{i}$: $\{e~|~e\in\mathcal{S}_i\land \forall i'\leqslant i, f_{i'}(e)\leqslant \frac{\lambda_{i'}}{2}\}$, the set of mice keys.
    
    \item $\mathcal{S}^{1}_{i}$:  $\{e~|~e\in\mathcal{S}_i\land \exists i'\leqslant i, f_{i'}(e) > \frac{\lambda_{i'}}{2}\}$, the set of elephant keys.
    
    \item $F_i$: $\sum_{\{e\in\mathcal{S}^{0}_i\}} f_i(e)$, the total frequency of mice keys in $\mathcal{S}^{0}_{i}$.
    
    \item $C_i$: $|\mathcal{S}^{1}_i|$, the number of elephant keys in $\mathcal{S}^{1}_{i}$.
    
    \item $\mathcal{S}^{0}_{i,j}$: $\{e~|~e\in\mathcal{S}^{0}_{i}\land h(e)=j\}$, the set of mice keys that are mapped to the $j$-th bucket.
    
    \item $\mathcal{S}^{1}_{i,j}$: $\{e~|~e\in\mathcal{S}^{1}_{i}\land h(e)=j\}$, the set of elephant keys that are mapped to the $j$-th bucket.
    
    \item $F_{i,j}$: $\sum_{\{e\in\mathcal{S}^{0}_{i,j}\}} f_i(e)$, the total frequency of mice keys in $\mathcal{S}^{0}_{i,j}$.
    
    \item $C_{i,j}$: $|\mathcal{S}^{1}_{i,j}|$, the number of elephant keys in $\mathcal{S}^{1}_{i,j}$.
    
    \item $\mathcal{P}_{i,k}$: $\{e_1,\cdots,e_{k}\}$, a subset of $\mathcal{S}_i$ composed of the first $k$ keys.
    
    \item $f^{P}_{i,k}$: $\sum_{\{e\in\mathcal{P}_{i,k-1}\cap\mathcal{S}^{0}_{i,h(e_k)}\}} f_i(e)$, the total frequency of mice keys with a smaller index that conflicts with key $e_k$.
    
    \item $c^{P}_{i,k}$: $\left|\{e~|~e\in\mathcal{P}_{i,k-1}\cap\mathcal{S}^{1}_{i,h(e_k)}\}\right|$, the number of elephant keys with a smaller index that conflicts with key $e_k$.
\end{enumerate}

\subsection{Properties in One Layer}
\label{app:filter}

This section aims to prove that only a small proportion of the keys inserted into the $i$-th layer will be inserted into the (i+1)-th layer. 

\begin{theorem}
    
\label{app:lem:x}    
    (Theorem \ref{Theorem:x:y})
    Let
    \begin{align*}
        X_{i,k}=
            \begin{cases}
              0 & C_{i, h(e_k)}=0 \land f^{P}_{i,k}\leqslant \frac{\lambda_i}{2}
            \\
              f_i(e_k) &  C_{i, h(e_k)}=0\land f^{P}_{i,k} > \frac{\lambda_i}{2}
            \\
              f_i(e_k) & C_{i, h(e_k)}>0
            \end{cases},
        &&
        X_i=\sum_{\{e_k\in\mathcal{S}^{0}_{i}\}} X_{i,k}.
    \end{align*}
    The total frequency of the mice keys in the $i$-th layer leaving it does not exceed $X_i$, \ie,
    \begin{align*}
        F_{i+1}
        \leqslant
        \sum_{\{e\in\mathcal{S}^{0}_i\cap\mathcal{S}_{i+1}\}} f_{i+1}(e) \leqslant X_{i}.
    \end{align*}
\end{theorem}

\begin{proof}
    For the mice keys in the $j$-th bucket of the $i$-th layer, let the number of times they leave be
    $F'_{i,j}=\sum_{\{e\in\mathcal{S}^{0}_{i,j}\cap\mathcal{S}^{0}_{i+1}\}} f_{i+1}(e).$
    Since a bucket can hold at least $\lambda_i$ packets of the key, we have:
    \begin{align*}
    \begin{cases}
        F'_{i,j}=0 & C_{i,j}=0 \land F_{i,j} \leqslant \lambda_i
        \\
        F'_{i,j}\leqslant F_{i,j}-\lambda_i & Cs_{i,j}=0 \land F_{i,j} > \lambda_i
        \\
        F'_{i,j}\leqslant F_{i,j} & C_{i,j}>0
    \end{cases}.
    \end{align*}
    When $C_{i,j}=0 \land F_{i,j} > \lambda_i$, exists $k'$ satisfies
    \begin{align*}
        \sum_{\{e_k\in\mathcal{S}^{0}_{i,j}\land k< k'\}} f_{i}(e_k)
        \leqslant \frac{\lambda_i}{2}
        \leqslant
        \sum_{\{e_k\in\mathcal{S}^{0}_{i,j}\land k\leqslant k'\}} f_{i}(e_k)
        \leqslant \lambda_i.
    \end{align*}
    Then for and only for any $e_k\in\mathcal{S}^{0}_{i,j}\land k\leqslant k'$, there is $X_{i,k}=0$, and
    \begin{align*}
        F'_{i,j}
        &\leqslant
        \left(
        \sum_{\{e_k\in\mathcal{S}^{0}_{i,j}\land k\leqslant k'\}} f_{i}(e_k)
        +
        \sum_{\{e_k\in\mathcal{S}^{0}_{i,j}\land k> k'\}} f_{i}(e_k)
        \right)
        -
        \lambda_i
        \\
        &\leqslant
        0 + \sum_{\{e_k\in\mathcal{S}^{0}_{i,j}\land k> k'\}} f_{i}(e_k)
        \\
        &\leqslant
        \sum_{\{e_k\in\mathcal{S}^{0}_{i,j}\land k\leqslant k'\}} X_{i, k}
        +
        \sum_{\{e_k\in\mathcal{S}^{0}_{i,j}\land k> k'\}} X_{i,k}.
    \end{align*}
    Then we have $F'_{i,j}\leqslant \sum_{\{e_k\in\mathcal{S}^{0}_{i,j}\}} X_{i,k}$, and
    \begin{align*}
        \sum_{\{e\in\mathcal{S}^{0}_i\cap\mathcal{S}_{i+1}\}} f_{i+1}(e)
        =
        \sum_{j=1}^{w_i} F'_{i,j}
        \leqslant
        \sum_{j=1}^{w_i} \sum_{\{e_k\in\mathcal{S}^{0}_{i,j}\}} X_{i,k}
        =
        X_{i}.
    \end{align*}

\end{proof}

Similarly, we have the following lemma.
\begin{theorem}
    
\label{app:lem:y}
  Let 
    \begin{align*}
        &Y_{i,k}=
            \begin{cases}
              0 &  c^{P}_{i,k}=0\land F_{i,h(e_k)}\leqslant \lambda_i,
            \\
              2 & c^{P}_{i,k}=0\land F_{i,h(e_k)}> \lambda_i
            \\
              2 & c^{P}_{i,k}>0.
            \end{cases},
        &&
        Y_{i}=\sum_{e_k\in\mathcal{S}^{1}_{i}} Y_{i,k}.
    \end{align*}
    The number of distinct elephant keys in the $i$-th layer leaving it does not exceed $Y_i$, \ie,
    \begin{align*}
        \left|\mathcal{S}^{1}_{i}\cap\mathcal{S}^{1}_{i+1}\right|\leqslant Y_{i}.
    \end{align*}
\end{theorem}

\begin{proof}
    For the elephant keys in the $j$-th bucket of the $i$-th layer, 
    $\sum_{\{e_k\in\mathcal{S}^{1}_{i,j}\}} Y_{i,k}<C_{i,j}$
    if and only if
    $C_{i,j}=1\land F_{i,j}\leqslant \lambda_i$.
    In this case, the number of collisions in the bucket does not exceed $\lambda_i$, and no key enters the $(i+1)$-th layer.
    Thus we have $|\mathcal{S}^{1}_{i,j}\cap\mathcal{S}^{1}_{i+1}|\leqslant \sum_{\{e_k\in\mathcal{S}^{1}_{i,j}\}} Y_{i,j}$, and
    \begin{align*}
        \left|\mathcal{S}^{1}_{i}\cap\mathcal{S}^{1}_{i+1}\right|
        =
        \sum_{j=1}^{w_i} |\mathcal{S}^{1}_{i,j}\cap\mathcal{S}^{1}_{i+1}|
        \leqslant
        \sum_{j=1}^{w_i} \sum_{\{e_k\in\mathcal{S}^{1}_{i,j}\}} Y_{i,j}
        =
        Y_{i}.
    \end{align*}
\end{proof}

\begin{theorem}
    
    Let
$W=\frac{4N(R_wR_\lambda)^{6}}{\Lambda(R_w-1)(R_\lambda-1)}$,
$\alpha_i= \frac{\Vert F\Vert_1}{(R_wR_\lambda)^{i-1}}$,
$\beta_i= \frac{\alpha_i}{\frac{\lambda_i}{2}}$,
$\gamma_i=(R_wR_\lambda)^{(2^{i-1}-1)}$, and
$p_i = (R_wR_\lambda)^{-(2^{i-1}+4)}$.
Under the conditions of
    $F_i\leqslant\frac{\alpha_i}{\gamma_i}$ and
    $C_i\leqslant\frac{\beta_i}{\gamma_i}$,
we have:
    \begin{align*}
        \Pr\left(X_{i,k}>0~|~X_{i,1},\cdots,X_{i,k-1}\right)
        \leqslant
        p_i,
        && \forall e_k\in\mathcal{S}^{0}_{i}.
        \\
        \Pr\left(Y_{i,k}>0~|~Y_{i,1},\cdots,Y_{i,k-1}\right)
        \leqslant 
        \frac{3}{4} p_i,
        && \forall e_k\in\mathcal{S}^{1}_{i}.
    \end{align*}
\label{app:theo:markov}
\end{theorem}

\begin{proof}
By using Markov's inequality, we have
\begin{align*}
    &
    \Pr\left(X_{i,k}>0~|~X_{i,1},\cdots,X_{i,k-1}\right)
    \\
    =&
    \Pr\left(
    \begin{aligned}
    \left(C_{i, h(e_k)}=0\land f^{P}_{i,k} > \frac{\lambda_i}{2}\right)
    \\\lor\quad
    C_{i, h(e_k)}>0
    \end{aligned}
    ~|~X_{i,1},\cdots,X_{i,k-1}
    \right)
    \\
    \leqslant&
    \quad
    \begin{aligned}
    \Pr\left(C_{i, h(e_k)}>0~|~X_{i,1},\cdots,X_{i,k-1}\right)
    \\+\quad
    \Pr\left(F_{i,h(e_k)} - f_i(e_k)> \frac{\lambda_i}{2}~|~X_{i,1},\cdots,X_{i,k-1}\right)
    \end{aligned}
    \\
    \leqslant&
    \quad
    \begin{aligned}
    \frac{E(C_{i, h(e_k)}~|~X_{i,1},\cdots,X_{i,k-1})}{1}
    \\ + \quad
    \frac{E(F_{i,h(e_k)}-f_i(e_k)~|~X_{i,1},\cdots,X_{i,k-1})}{\frac{\lambda_i}{2}}
    \end{aligned}
    \\
    \leqslant&
    \frac{C_i}{w_i} + \frac{2F_i}{\lambda_i w_i}
\end{align*}
\begin{align*}
    &
    \Pr\left(Y_{i,k}>0~|~Y_{i,1},\cdots,Y_{i,k-1}\right)
    \\
    =&
    \Pr\left(\left(c^{P}_{i,k}=0\land F_{i,h(e_k)}> \lambda_i\right)\lor c^{P}_{i,k}>0~|~Y_{i,1},\cdots,Y_{i,k-1}\right)
    \\
    \leqslant&\quad
    \begin{aligned}
    \Pr\left(C_{i, h(e_k)}-1>0~|~Y_{i,1},\cdots,Y_{i,k-1}\right)
    \\+\quad
    \Pr\left(F_{i,h(e_k)} > \lambda_i~|~Y_{i,1},\cdots,Y_{i,k-1}\right)
    \end{aligned}
    \\
    \leqslant&\quad
    \begin{aligned}
    \frac{E(C_{i, h(e_k)}-1~|~Y_{i,1},\cdots,Y_{i,k-1})}{1}
    \\+\quad
    \frac{E(F_{i,h(e_k)}~|~Y_{i,1},\cdots,Y_{i,k-1})}{\lambda_i}
    \end{aligned}
    \\
    \leqslant&
    \frac{C_i}{w_i} + \frac{F_i}{\lambda_i w_i}.
\end{align*}
Recall that
$w_i=\lceil\frac{W (R_{w}-1)}{R_{w}^i}\rceil$ and
$\lambda_i = \frac{\Lambda (R_\lambda-1)}{R_\lambda^i}$
, under the conditions of
    $F_i\leqslant\frac{\alpha_i}{\gamma_i}$ and
    $C_i\leqslant\frac{\beta_i}{\gamma_i}$,
we have
\begin{align*}
    &
    \Pr\left(X_{i,k}>0~|~X_{i,1},\cdots,X_{i,k-1}\right)
    \\
    \leqslant&
    \frac{\beta_i}{\gamma_i w_i} + \frac{2\alpha_i}{\gamma_i \lambda_i w_i}
    =
    \frac{4\alpha_i}{\gamma_i \lambda_i w_i}
    \leqslant
    \frac{1}{(R_wR_\lambda)^{2^{i-1}+4}}=p_i.
    \\
    &
    \Pr\left(Y_{i,k}>0~|~Y_{i,1},\cdots,Y_{i,k-1}\right)
    \\
    \leqslant&
    \frac{\beta_i}{\gamma_i w_i} + \frac{\alpha_i}{\gamma_i \lambda_i w_i}
    =
    \frac{3\alpha_i}{\gamma_i \lambda_i w_i}
    \leqslant
    \frac{3}{4(R_wR_\lambda)^{2^{i-1}+4}}\leqslant \frac{3}{4}p_i.
\end{align*}
\end{proof}

\begin{theorem}
    (Theorem \ref{app:theo:exp_1})
Let
$W=\frac{4N(R_wR_\lambda)^{6}}{\Lambda(R_w-1)(R_\lambda-1)}$,
$\alpha_i= \frac{\Vert F\Vert_1}{(R_wR_\lambda)^{i-1}}$,
$\beta_i= \frac{\alpha_i}{\frac{\lambda_i}{2}}$,
$\gamma_i=(R_wR_\lambda)^{(2^{i-1}-1)}$, and
$p_i = (R_wR_\lambda)^{-(2^{i-1}+4)}$.
Under the conditions of
    $F_i\leqslant\frac{\alpha_i}{\gamma_i}$ and
    $C_i\leqslant\frac{\beta_i}{\gamma_i}$,
we have
\begin{align*}
    \Pr\left(X_{i}>(1+\Delta)\frac{p_i\alpha_i}{\gamma_i}\right)
    \leqslant
    \exp\left(-(\Delta-(e-2))\frac{2p_i\alpha_i}{\lambda_i\gamma_i}\right).
\end{align*}
and
\begin{align*}
    \Pr\left(Y_{i}>(1+\Delta)\frac{3}{2}\frac{p_i\beta_i}{\gamma_i}\right)
    \leqslant
    \exp\left(-(\Delta-(e-2))\frac{3p_i\beta_i}{4\gamma_i}\right).
\end{align*}
\end{theorem}

\begin{proof}
According to Theorem \ref{app:theo:markov},
\begin{align*}
    &
    \Pr\left(
    \frac{X_{i,k}}{\frac{\lambda_i}{2}}=\frac{f_i(e_k)}{\frac{\lambda_i}{2}}
    ~|~
    \frac{X_{i,1}}{\frac{\lambda_i}{2}},\cdots,\frac{X_{i,k-1}}{\frac{\lambda_i}{2}}
    \right)
    \leqslant p_i.
    \\
    &
    \Pr\left(\frac{Y_{i,k}}{2}=1
    ~|~
    \frac{Y_{i,1}}{2}, \cdots, \frac{Y_{i,k-1}}{2}
    \right)
    \leqslant \frac{3}{4}p_i.
\end{align*}
According to Lemma \ref{app:basic},
\begin{align*}
    &
    \Pr\left(
    X_i>(1+\Delta)\frac{p_i\alpha_i}{\gamma_i}
    \right)
    \leqslant
    \Pr\left(
    X_i>(1+\Delta)p_iF_i
    ~|~F_i=\frac{\alpha_i}{\gamma_i}
    \right)
    \\
    =&
    \Pr\left(
    \sum_{\{e_k\in\mathcal{S}^{0}_{i}\}} \frac{X_{i,k}}{\frac{\lambda_i}{2}}>(1+\Delta)p_i\sum_{\{e_k\in\mathcal{S}^{0}_{i}\}}\frac{f_i(e_k)}{\frac{\lambda_i}{2}}
    ~|~F_i=\frac{\alpha_i}{\gamma_i}
    \right)
    \\
    \leqslant&
    \exp\left(-(\Delta-(e-2))\frac{\alpha_i}{\gamma_i\frac{\lambda_i}{2}}p_i\right)
    =
    \exp\left(-(\Delta-(e-2))\frac{2p_i\alpha_i}{\lambda_i\gamma_i}\right).
    \\
    &
    \Pr\left(
    Y_i>(1+\Delta)\frac{3}{2}\frac{p_i\beta_i}{\gamma_i}
    \right)
    \leqslant
    \Pr\left(
    Y_i>(1+\Delta)\frac{3}{2}p_iC_i
    ~|~ C_i=\frac{\beta_i}{\gamma_i}
    \right)
    \\
    =&
    \Pr\left(
    \sum_{\{e_k\in\mathcal{S}^{1}_{i}\}} \frac{Y_{i,k}}{2}>(1+\Delta)\frac{3}{4} p_i\sum_{\{e_k\in\mathcal{S}^{1}_{i}\}}\frac{2}{2}
    ~|~ C_i=\frac{\beta_i}{\gamma_i}
    \right)
    \\
    \leqslant&
    \exp\left(-(\Delta-(e-2))\frac{\beta_i}{\gamma_i}\frac{3}{4}p_i\right)
    =
    \exp\left(-(\Delta-(e-2))\frac{3p_i\beta_i}{4\gamma_i}\right).
\end{align*}

\end{proof}

\begin{theorem}
\label{app:theo:gamma}
(Theorem \ref{theo:gamma})
Let
$R_wR_\lambda\geqslant2$,
$W=\frac{4N(R_wR_\lambda)^{6}}{\Lambda(R_w-1)(R_\lambda-1)}$,
$\alpha_i= \frac{\Vert F\Vert_1}{(R_wR_\lambda)^{i-1}}$,
$\beta_i= \frac{\alpha_i}{\frac{\lambda_i}{2}}$,
$\gamma_i=(R_wR_\lambda)^{(2^{i-1}-1)}$, and
$p_i = (R_wR_\lambda)^{-(2^{i-1}+4)}$.
We have
\begin{align*}
&    
    \Pr\left(F_{i+1}>\frac{\alpha_{i+1}}{\gamma_{i+1}}
    ~|~
    F_{i}\leqslant\frac{\alpha_{i}}{\gamma_{i}} \land C_{i}\leqslant\frac{\beta_{i}}{\gamma_{i}}\right)
    \\
    \leqslant&
    \exp\left(-(9-e)\frac{2p_i\alpha_i}{\lambda_i\gamma_i}\right).
    \\
&
    \Pr\left(C_{i+1}>\frac{\beta_{i+1}}{\gamma_{i+1}}
    ~|~
    F_{i}\leqslant\frac{\alpha_{i}}{\gamma_{i}} \land
    C_{i}\leqslant\frac{\beta_{i}}{\gamma_{i}}\right)
    \\
    \leqslant&
    \exp\left(-(5-e)\frac{2p_i\alpha_i}{\lambda_i\gamma_i}\right)
    +
    \exp\left(-(\frac{11}{3}-e)\frac{3p_i\beta_i}{4\gamma_i}\right).
\end{align*}
\end{theorem}

\begin{proof}
According to settings, we have
\begin{align*}
    p_i\frac{\alpha_i}{\gamma_i}
    &=
    \frac{\Vert F\Vert_1}{(R_wR_\lambda)^{(2^{i}+i+2)}}
    \leqslant
    \frac{1}{8}\frac{\alpha_{i+1}}{\gamma_{i+1}}
    \\
    p_i\frac{\beta_i}{\gamma_i}
    &=
    p_i\frac{\alpha_i}{\gamma_i\frac{\lambda_i}{2}}
    \leqslant
    \frac{1}{8}\frac{\alpha_{i+1}}{\gamma_{i+1}\frac{\lambda_{i+1}}{2}}
    =
    \frac{1}{8}\frac{\beta_{i+1}}{\gamma_{i+1}}.
\end{align*}
Recall that $C_{i+1}=|\mathcal{S}^{1}_{i+1}\cap\mathcal{S}^{0}_i|+|\mathcal{S}^{1}_{i+1}\cap\mathcal{S}^{1}_i|$, and
\begin{align*}
|\mathcal{S}^{1}_{i+1}\cap\mathcal{S}^{0}_i|
\leqslant
\frac{\sum_{\{e\in\mathcal{S}^{0}_i\cap\mathcal{S}_{i+1}\}} f_i(e)}{\frac{\lambda_{i+1}}{2}}
\leqslant
\frac{X_i}{\frac{\lambda_{i+1}}{2}}
\end{align*}
Let $\Gamma_i=\left(F_{i}\leqslant\frac{\alpha_{i}}{\gamma_{i}} \land
    C_{i}\leqslant\frac{\beta_{i}}{\gamma_{i}}\right)$, according to Theorem \ref{app:lem:x} and Theorem \ref{app:theo:exp_1}, 
\begin{align*}
    \Pr(F_{i+1}>\frac{\alpha_{i+1}}{\gamma_{i+1}}
    ~|~\Gamma_i)
    & \leqslant
    \Pr\left(X_{i}>8 p_i\frac{\alpha_i}{\gamma_i} ~|~\Gamma_i\right)\\
    & \leqslant
    \exp\left(-(9-e)\frac{2p_i\alpha_i}{\lambda_i\gamma_i}\right).
\end{align*}
According to Theorem \ref{app:lem:y} and Theorem \ref{app:theo:exp_1}, 
\begin{align*}
    &
    \Pr(C_{i+1}>\frac{\beta_{i+1}}{\gamma_{i+1}}~|~\Gamma_i)
    \\
    =&
    \Pr(|\mathcal{S}^{1}_{i+1}\cap\mathcal{S}^{0}_i|+|\mathcal{S}^{1}_{i+1}\cap\mathcal{S}^{1}_i|>\frac{\beta_{i+1}}{\gamma_{i+1}}~|~\Gamma_i)
    \\
    \leqslant&
    \Pr(|\mathcal{S}^{1}_{i+1}\cap\mathcal{S}^{0}_i|>\frac{\beta_{i+1}}{2\gamma_{i+1}}
    \lor
    |\mathcal{S}^{1}_{i+1}\cap\mathcal{S}^{1}_i|>\frac{\beta_{i+1}}{2\gamma_{i+1}}~|~\Gamma_i)
    \\
    \leqslant&
    \Pr(\frac{X_i}{\frac{\lambda_{i+1}}{2}}>\frac{\beta_{i+1}}{2\gamma_{i+1}}~|~\Gamma_i)
    +
    \Pr(Y_i>\frac{\beta_{i+1}}{2\gamma_{i+1}}~|~\Gamma_i)
    \\
    \leqslant&
    \Pr(X_i > 4p_i\frac{\alpha_i}{\gamma_i}~|~\Gamma_i)
    +
    \Pr(Y_i>4p_i\frac{\beta_i}{\gamma_i}
    ~|~\Gamma_i)
    \\
    \leqslant&
    \exp\left(-(5-e)\frac{2p_i\alpha_i}{\lambda_i\gamma_i}\right)
    +
    \exp\left(-(\frac{11}{3}-e)\frac{3p_i\beta_i}{4\gamma_i}\right)
    .
\end{align*}
\end{proof}

\subsection{Space and Time Complexity}
\label{app:finalcon}

\begin{theorem}
\label{app:theo:main}
(Theorem \ref{theo:main})
Let
$R_wR_\lambda\geqslant 2$,
$W=\frac{4N(R_wR_\lambda)^{6}}{\Lambda(R_w-1)(R_\lambda-1)}$,
$\alpha_i= \frac{\Vert F\Vert_1}{(R_wR_\lambda)^{i-1}}$,
$\beta_i= \frac{\alpha_i}{\frac{\lambda_i}{2}}$,
$\gamma_i=(R_wR_\lambda)^{(2^{i-1}-1)}$, and
$p_i = (R_wR_\lambda)^{-(2^{i-1}+4)}$.
For given $\Lambda$ and $\Delta<\frac{1}{4}$, 
let $d$ be the root of the following equation
\begin{align*}
    \frac{R_\lambda^{d}}{(R_wR_\lambda)^{(2^{d}+d)}}=\Delta_1\frac{\Lambda}{N}\ln(\frac{1}{\Delta}).
\end{align*}
And use an SpaceSaving of size $\Delta_2\ln(\frac{1}{\Delta})$ (as the $(d+1)$-layer), then 
\begin{align*}
    \Pr\left(\forall \textit{ item } e, \left|\hat{f}(e)-f(e)\right|\leqslant\Lambda\right)\geqslant 1-\Delta,
\end{align*}
where
\begin{align*}
    \Delta_1=2R_w^2R_\lambda^2(R_\lambda-1),
    &&
    \Delta_2=3\left(\frac{R_wR_\lambda^2}{R_\lambda-1}\right)
    \Delta_1=6R_w^3R_\lambda^4.
\end{align*}
\end{theorem}

\begin{proof}
Recall that 
$\Gamma_i=\left(F_{i}\leqslant\frac{\alpha_{i}}{\gamma_{i}} \land C_{i}\leqslant\frac{\beta_{i}}{\gamma_{i}}\right)$,
When all conditions $\Gamma_i$ (including $\Gamma_{d+1}$) are true, we have
\begin{align*}
    C_{d+1}
    \leqslant&
    \frac{\beta_{d+1}}{\gamma_{d+1}}
    =
    \frac{2 N R_\lambda^{d+1}}{(R_wR_\lambda)^{(2^{d}+d-1)}(R_\lambda-1)\Lambda}
    =
    \left(\frac{2R_wR_\lambda^2}{R_\lambda-1}\right)
    \Delta_1\ln(\frac{1}{\Delta}).
    \\
    F_{d+1}
    \leqslant&
    \frac{\alpha_{d+1}}{\gamma_{d+1}}
    =
    \frac{\lambda_{d+1}}{2}\frac{\beta_{d+1}}{\gamma_{d+1}}
    =
    \lambda_{d+1}\left(\frac{R_wR_\lambda^2}{R_\lambda-1}\right)
    \Delta_1\ln(\frac{1}{\Delta}).
\end{align*}
Since we use an SpaceSaving of size $\Delta_2\ln(\frac{1}{\Delta})>C_{d+1}$, it can record all elephant keys without error, and the estimation error for mice keys does not exceed
\begin{align*}
    \frac{F_{d+1}}{\Delta_2\ln(\frac{1}{\Delta})-C_{d+1}}
    \leqslant
    \frac{\lambda_{d+1}\left(\frac{R_wR_\lambda^2}{R_\lambda-1}\right)
    \Delta_1\ln(\frac{1}{\Delta})}{\Delta_2\ln(\frac{1}{\Delta})-\left(\frac{2R_wR_\lambda^2}{R_\lambda-1}\right)
    \Delta_1\ln(\frac{1}{\Delta})}
    =\lambda_{d+1}
\end{align*}
Therefore, for any item $e$,
\begin{align*}
    \left|\hat{f}(e)-f(e)\right|
    =
    \sum_{i=1}^{d} \lambda_i
    \leqslant
    \sum_{i=1}^{\infty} \frac{\Lambda(R_\lambda -1)}{R_\lambda^{i}}
    =
    \Lambda
\end{align*}
Next, we deduce the probability that at least one condition $\Gamma_i$ is false.
Note that
\begin{align*}
    \left.
    \begin{aligned}
    & (\frac{11}{3}-e)\frac{3p_i\beta_i}{4\gamma_i}
    \\
    & (9-e)\frac{2p_i\alpha_i}{\lambda_i\gamma_i}
    \\
    & (5-e)\frac{2p_i\alpha_i}{\lambda_i\gamma_i}
    \end{aligned}
    \right\}
    \geqslant\frac{p_i\alpha_i}{\lambda_i\gamma_i}.
\end{align*}
Then According to Theorem \ref{app:theo:gamma}, we have
\begin{align*}
    &
    \Pr\left(\lnot\left(\bigwedge_{i=1}^{d}\Gamma_{i+1}\right)\right)
    =
    \Pr\left(\bigvee_{i=1}^{d} \lnot\Gamma_{i+1}\right)
    =
    \Pr\left(\bigvee_{i=1}^{d}\left(\bigwedge_{j=1}^{i}\Gamma_j\land\lnot\Gamma_{i+1}\right)\right)
    \\
    \leqslant&
    \sum_{i=1}^{d}\Pr\left(\Gamma_{i}\land\lnot\Gamma_{i+1}\right)
    \leqslant
    \sum_{i=1}^{d}\Pr\left(\lnot\Gamma_{i+1}~|~\Gamma_{i}\right)
    \\
    \leqslant&
    \sum_{i=1}^{d}
    \left(
    \begin{aligned}
    \exp\left(-(\frac{11}{3}-e)\frac{3p_i\beta_i}{4\gamma_i}\right)
    \\+
    \exp\left(-(9-e)\frac{2p_i\alpha_i}{\lambda_i\gamma_i}\right)
    +
    \exp\left(-(5-e)\frac{2p_i\alpha_i}{\lambda_i\gamma_i}\right)
    \end{aligned}
    \right)
    \\
    \leqslant&
    \sum_{i=1}^{d} 3\exp\left(-\frac{p_i\alpha_i}{\lambda_i\gamma_i}\right).
\end{align*}
Note that
\begin{align*}
    \exp\left(-\frac{p_d\alpha_d}{\lambda_d\gamma_d}\right)
    =&
    \exp\left(-\frac{N R_\lambda^{d}}{(R_wR_\lambda)^{(2^{d}+d+2)}\Lambda(R_\lambda-1)}\right)
    \\
    =&
    \exp\left(-\frac{1}{R_w^2R_\lambda^2(R_\lambda-1)}\Delta_1\ln(\frac{1}{\Delta})\right)
    \\
    =&
    \Delta^{\left(\frac{1}{R_w^2R_\lambda^2(R_\lambda-1)}\Delta_1\right)}
    =\Delta^{2}.
\end{align*}
Since $\Delta\leqslant1$, and the monotonicty of $\exp\left(-\frac{p_d\alpha_d}{\lambda_d\gamma_d}\right)$, we have
\begin{align*}
    \exp\left(-\frac{p_i\alpha_i}{\lambda_i\gamma_i}\right)
    =&
    \exp\left(-\frac{p_{i+1}\alpha_{i+1}}{\lambda_{i+1}\gamma_{i+1}}\cdot R_w^{(2^{i}+1)}R_\lambda^{(2^{i})}\right)
    \\
    \leqslant&
    \exp\left(-\frac{p_{i+1}\alpha_{i+1}}{\lambda_{i+1}\gamma_{i+1}}\right)^{R_wR_\lambda}
    \leqslant
    \exp\left(-\frac{p_{i+1}\alpha_{i+1}}{\lambda_{i+1}\gamma_{i+1}}\right)^{2}
    \\
    \leqslant&
    \Delta^2\exp\left(-\frac{p_{i+1}\alpha_{i+1}}{\lambda_{i+1}\gamma_{i+1}}\right)
\end{align*}
Therefore, we have
\begin{align*}
    \sum_{i=1}^{d} 3\exp\left(-\frac{p_i\alpha_i}{\lambda_i\gamma_i}\right)
    \leqslant
    3\sum_{i=1}^{d} \Delta^{2i}
    \leqslant
    \left(\frac{3\Delta}{1-\Delta^2}\right)\Delta
    \leqslant\Delta.
\end{align*}
In other words,
\begin{align*}
    \Pr\left(\forall \textit{ item } e, \left|\hat{f}(e)-f(e)\right|\leqslant\Lambda\right)\geqslant 1-\Delta,
\end{align*}
which leads to a weaker conclusion,
\begin{align*}
    \forall \textit{ item } e, \Pr\left(\left|\hat{f}(e)-f(e)\right|\leqslant\Lambda\right)\geqslant 1-\Delta.
\end{align*}
\end{proof}

\begin{theorem}
Using the same settings as Theorem \ref{app:theo:main}, the space complexity of the algorithm is $O(\frac{N}{\Lambda}+\ln(\frac{1}{\Delta}))$, and the time complexity of the algorithm is amortized $O(1+\Delta\ln\ln(\frac{N}{\Lambda}))$.
\end{theorem}

\begin{proof}
Recall that $d$ is the root of the equation
\begin{align*}
    \frac{R_\lambda^{d}}{(R_wR_\lambda)^{(2^{d}+d)}}=\Delta_1\frac{\Lambda}{N}\ln(\frac{1}{\Delta}),
\end{align*}
which means $d=O\left(\ln\ln(\frac{N}{\Lambda})\right)$.
Therefore, total space used by the data structure is
\begin{align*}
    \sum_{i=1}^{d} w_i+\Delta_1\ln(\frac{1}{\Delta})
    =&
    \sum_{i=1}^{d} \lceil\frac{W (R_{w}-1)}{R_{w}^i}\rceil
    +O(\ln(\frac{1}{\Delta}))
    \\
    \leqslant&
    \frac{4N(R_wR_\lambda)^{6}}{\Lambda(R_w-1)(R_\lambda-1)}+d+O(\ln(\frac{1}{\Delta}))
    \\
    =&
    O(\frac{N}{\Lambda}+\ln(\frac{1}{\Delta}))
\end{align*}
Next, we analyze the time complexity.
When all condition $\Gamma_i$ are true, for a new item $e\not\in\mathcal{S}_{1}$, the probability that it enters the $(i+1)$-th layer from the $i$-th layer is
$\frac{\frac{F_i}{\frac{\lambda_2}{2}}+C_i}{w_i}\leqslant p_i$.
Thus the time complexity of insert item $e$ does noes exceed
\begin{align*}
    (1-\Delta)\cdot(1+\sum_{i=1}^{d} p_i)
    +
    \Delta\cdot d
    =
    O(1+\Delta\ln\ln(\frac{N}{\Lambda})).
\end{align*}
%
\end{proof}	

\end{document}